\documentclass[fleqn,usenatbib]{mnras}

\usepackage{newtxtext,newtxmath}

\usepackage[T1]{fontenc}

\DeclareRobustCommand{\VAN}[3]{#2}
\let\VANthebibliography\thebibliography
\def\thebibliography{\DeclareRobustCommand{\VAN}[3]{##3}\VANthebibliography}

\usepackage{graphicx}	
\usepackage{amsmath}	

\title[GRBs with RINGO3]{Polarimetry and Photometry of Gamma-Ray Bursts Afterglows with RINGO3}

\author[M. Shrestha et al.]{M. Shrestha,$^{1}$\thanks{E-mail: m.shrestha@ljmu.ac.uk}
I. A. Steele,$^{1}$
S. Kobayashi,$^{1}$
R J. Smith,$^{1}$
C. Guidorzi, $^{2,3,4}$
N. Jordana-Mitjans, $^{5}$
\newauthor H. Jermak,$^{1}$
D. Arnold,$^{1}$
C. G. Mundell,$^{5}$
A. Gomboc $^{6}$
\\
$^{1}$Astrophysics Research Institute, Liverpool John Moores University, Liverpool Science Park, 146 Brownlow Hill, UK \\
Liverpool L3 5RF, UK\\
$^{2}$Department of Physics and Earth Science, University of Ferrara, via Saragat 1, 44122 Ferrara, Italy \\
$^{3}$ INFN -- Sezione di Ferrara, via Saragat 1, I 44122, Ferrara, Italy\\
$^{4}$ INAF -- Osservatorio di Astrofisica e Scienza dello Spazio di Bologna, Via Piero Gobetti 101, I 40129 Bologna, Italy\\
$^{5}$Department of Physics, University of Bath, Claverton Down, Bath, BA2 7AY, UK\\
$^{6}$Center for Astrophysics and Cosmology, University of Nova Gorica, Vipavska 13, 5000 Nova Gorica, Slovenia \\}

\date{Accepted XXX. Received YYY; in original form ZZZ}

\pubyear{2021}

\begin{document}
\label{firstpage}
\pagerange{\pageref{firstpage}--\pageref{lastpage}}
\maketitle

\begin{abstract}
We present photometric and polarimetric measurements of gamma-ray burst (GRB) optical afterglows observed by the RINGO3 imaging polarimeter over its $\sim$7 year lifetime mounted on the Liverpool Telescope. During this time, RINGO3 responded to 67 GRB alerts. Of these, 28 had optical afterglows and a further ten were sufficiently bright for photometric and polarimetric analysis ($R\lessapprox{17}$). We present high quality multicolour light curves of ten sources: GRB 130606A, GRB 130610A, GRB 130612A, GRB 140430A, GRB 141220A, GRB 151215A, GRB 180325A, GRB 180618A, GRB 190114C, and GRB 191016A and polarimetry for seven of these (excluding GRB 130606A, GRB 130610A, and GRB 130612A, which were observed before the polarimetry mode was fully commissioned). Eight of these ten GRBs are classical long GRBs, one sits at the short-long duration interface with a $T_{90}$ $\sim$ 4 seconds and one is a classical short, hard burst with extended emission. We detect polarization for GRB 190114C and GRB 191016A. While detailed analyses of several of these GRBs have been published previously, here we present a uniform re-reduction and analysis of the whole sample and investigation of the population in a broad context relative to the current literature.  We use survival analysis to fully include the polarization upper limits in the comparison with other GRB properties, such as temporal decay rate, isotropic energy and redshift. We find no clear correlation
between polarization properties and wider sample properties and conclude that larger samples of early time polarimetry of GRB afterglows are required to fully understand GRB magnetic fields.
\end{abstract}




\section{Introduction}
Gamma-ray bursts (GRBs) are extremely energetic transients occurring at cosmological distances. They have been observed at various wavelengths ranging from gamma-rays to radio and these observations help us to piece together the puzzle of GRB physics. One of the proposed scenarios is that accretion onto a compact object powers the relativistic outflow and other prediction is the magnetar spindown as the powering source for the GRB \citep{Metzger_2011}. Internal dissipation in the outflow causes the prompt gamma-rays, and external shocks (interaction of the jet with local ambient medium) produce afterglow emission at various frequencies ranging from x-ray to radio \citep{Piran_1999, Zhang_2004}.

One of the most puzzling aspects of GRBs is the magnetic field properties of their jets which can shed light on the driving mechanism of the explosion \citep{Lazzati_2006, Toma_2013,Covino_2016,Kobayashi_2019}. Since GRBs are cosmological in nature, we cannot obtain the spatial resolution to understand the magnetic field strength and structure of the jets with most observational methods. Photometric observations of the forward and reverse shock afterglows can constrain the relative strength of the magnetic fields in the two shock regions, whereas polarimetry can provide information about the structure of the magnetic field in the original ejecta from the GRB central engine. Thus, polarization studies of early afterglows (few minutes after the burst) are an important technique to better understand the magnetic field properties of GRB jets. 

Most polarimetric studies have focused on the observation of polarization near the jet break to understand the jet opening angle \citep{Ghisellini_1999, Sari_1999, Rossi_2004}. These jet breaks occur $\sim$ 1 day after the burst, at times when we observe mostly the forward shock emission which is not highly polarized \citep{steele-ringo2-2017, kopac_2015, Jordana_2020}. Increased availability of robotic telescopes and instruments designed to observe polarization of transients have facilitated the observation of earlier time signals of these GRBs \citep{Steele_2004} (minutes to hours after the burst). The early afterglow includes the signatures of reverse shocks and the magnetic field structure of the jet itself; these reverse shock afterglows have higher levels of polarization compared to forward shocks \citep{Steele_2009,Mundell_2013,steele-ringo2-2017,Shrestha_2021}. These data can be utilized with current theoretical models to narrow down the physics of GRB jets. 

The Liverpool telescope (LT) which is a 2.0 meter fully autonomous robotic telescope at Observatorio del Roque de los Muchachos, La Palma \citep{Steele_2004}. LT is equipped with polarimeters designed for rapid observations. RINGO (2006 - 2009), RINGO2 (2010 -  2012), and RINGO3 (2013 - 2020) are a series of polarimeters that used a rapidly rotating polaroid analyser \citep{Jermak_2016,Arnold_2017} to observe rapidly fading sources with high accuracy. RINGO and RINGO2 observed very high polarization signals of early-time optical polarization in some GRBs \citep{Steele_2009, Mundell_2013}. All the RINGO2 GRB observations are presented in \citet{steele-ringo2-2017}. In this paper we present a unified analysis of all of the GRBs observed by final generation of the RINGO polarimeters, RINGO3. This increases the sample size of polarization data which will help us better understand the magnetic properties of GRB jets. 

In this paper, we present the results of the complete set of GRBs observed by RINGO3. We present photometric analysis of ten GRBs and polarimetric analysis of seven GRBs. The paper is arranged as follows; in Section \ref{sec:obs}, we present the design of RINGO3 and different observations performed during its run time. We describe the data reduction process in Section~\ref{sec:reduction} and present the polarimetric and photometric results in Section~\ref{sec:results}. We discuss the implications of these observations in Section~\ref{sec:discussions}. Finally, we provide concluding remarks in Section~\ref{sec:conclusions}.

\section{Instrument and Observations}\label{sec:obs}
In this paper we present observational data from three different instruments: RINGO3 (polarimeter), IO:O (imager), and RATCam (imager) on board the LT. Since it is a fully robotic telescope, it is optimal for time-domain astrophysics including GRB studies \citep{Guidorzi_2006}. For all the instruments, basic CCD reductions such as bias subtraction, dark subtraction, flat fielding and World Coordinate system fitting is done via an internal common pipeline \footnote{https://telescope.livjm.ac.uk/TelInst/Pipelines/}.

\subsection{RATCam and IO:O Imaging Cameras}

RATCam\footnote{https://telescope.livjm.ac.uk/TelInst/Inst/RATCam/} \citep{steele_2001} and IO:O \footnote{http://telescope.livjm.ac.uk/TelInst/Inst/IOO/}  were used for photometric observations. RATCam (field of view $4.6\times4.6$ arcmin) and IO:O ($10\times10$ arcmin) are optical CCD cameras equipped with $u'g'r'i'z'$ filters. In this paper, we present results from camera using the $r'$ filter because the wavelength range of this filter is closest to the $R$ band data of RINGO3 thus we can make a better comparison to the rest of the data set.

\subsection{RINGO3 polarimeter}\label{subsec:inst-ringo3}
RINGO3 \citep{Arnold_2012} was the third generation of fast-readout optical imaging polarimeters on board the LT and was observing from early 2013 to January 2020. It had a field of view of $4 \times 4$ arcmin, and used a polaroid that rotated at $\sim 0.4$-Hz. The instrument was designed using three separate electron multiplying CCDs to simultaneously observe polarized images in three different wavebands. The three wavebands have wavelength ranges of $7700-10000 $ \AA, $6500-7600$ \AA, and $3500-6400$  \AA. We convert these filters roughly corresponding to the standard astronomical $I$, (with $\lambda_{\rm eff} \sim 8500  $ \AA), $R$ ($\lambda_{\rm eff} \sim 7050 $ \AA ), 
and $V$ ($\lambda_{\rm eff} \sim 5300 $ \AA ) bands. Each camera obtained eight exposures per rotation which were synchronised with the phase of the polaroid's rotation. RINGO3 produced 24 CCD frames (8 per camera) every 2.3 seconds which were stacked per camera into 1 minute and 10 minutes blocks for each eight rotor position image. Data from these eight exposures were utilized to deduce linear Stokes vectors; explained in detail in Section~\ref{sec:reduction}. For 10 minute stacked data, we obtain a polarization accuracy up to $2.5\%$, $1.5\%$, and $0.5 \%$ for a 17 mag source in $I$, $R$, and $V$ filters respectively. Thus, we create a 17 mag cut off for robust polarimetric analysis. 

\subsection{Observations}\label{subsec:inst-obs}
\begin{figure*}
\centering
\includegraphics[width=2\columnwidth]{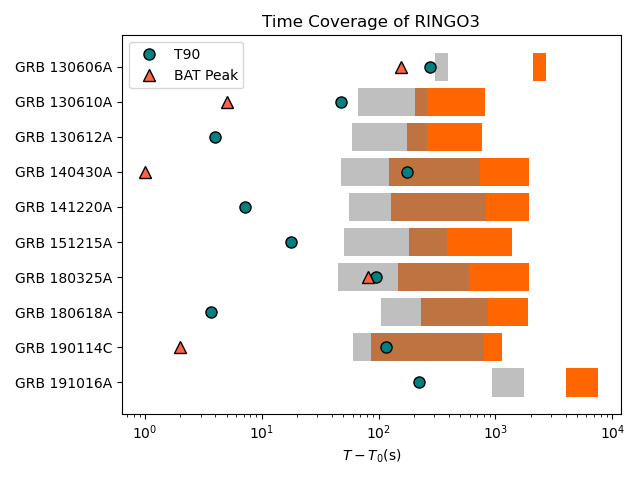}
\caption{ Plot of observation time coverage of all the GRBs analyzed in this paper where the orange bar represents the observer's time range, and the grey bar represents the time-dilation corrected range. Triangular points are peak time for BAT observation in observer's frame (we do not plot BAT peak time of 0 or less )  and T90 is presented as circle in observer's frame. }
\label{fig:time_coverage}
\end{figure*}

Between 2013 and 2020, a total of 67 GRB alerts as shown in Table~\ref{tab:alert} were observed by RINGO3 and 28 of them had optical counterpart. Out of 28 GRBs with optical counterparts, two had only one data point so they were excluded from this analysis. Three observations experienced instrumental issues and had incorrect pointing. Thirteen were too faint, with an $R$ magnitude greater than 17, to attempt RINGO3 photometric and polarimetric analysis. Thus, ten alerts had optical afterglows which were bright enough to perform RINGO3 photometry and polarimetry. Figure~\ref{fig:time_coverage} shows the observational time coverage of these GRBs in the observer's and time-dilation corrected time range, along with $T_{90}$ which is the duration between $5\%$ to $95\%$ of counts is measured and the Burst Alert Telescope (BAT) peak time in the observer's reference frame. Three of these afterglows were observed before December 2013; during that time period there were problems constraining the instrumental polarization induced by the two dichroic mirrors. Thus, we can only perform photometric analysis of these GRBs. In this paper, we present polarimetric results for seven out of ten bright afterglows observed by RINGO3. The properties of GRBs analysed in this paper are presented in Table~\ref{tab:general} which contains names of the GRBs, RA, DEC, RINGO3 observation duration, $T_{90}$, Galactic extinction, redshift, and related references. 

Results for GRB 140430A, GRB 141220A, GRB 190114C, and GRB 191016A have already been published separately in \cite{kopac_2015, Jordana_2021,Jordana_2020, Shrestha_2021} respectively and a detailed analysis of GRB~180618A is submitted for publication (Jordana-Mitjans et al, 2022, submitted).  In this paper we re-analyse these bursts as well as the data on the other unpublished events in order to allow a more homogeneous analysis of the entire sample.

\begin{table*}
    \center
     \caption{Properties of all the triggers observed by RINGO3 in $\sim$ 7 years. OT stands for optical transient. GCN stands for gamma-ray burst coordinates network which is a system that distributes the information about GRBs. }
     \resizebox{!}{.95\height}{\begin{tabular}{c|c|c|c|c|c|c|c}
   
          \hline
           \hline
         GRB & RA ($\degr$) & DEC ($\degr$) & OT  &  $T - T_0$ (s) & IO/RATCAM & LT GCN &  Note \\
          \hline
         130216A& 67.90 & 14.67 & NO & 678  & YES &- &- \\
         130328A & - & - & - & 246 &  - & - &No GCN \\
         130408A & 134.40 & -32.36 & YES &  517 & YES & 14362 &Only one observation \\
         130427A & 173.13 & 27.69 & YES & 47158 & YES  & - &Limited number of observation\\
         130504A &272.45 & -16.31 & NO & 193 & YES &- &Only upper limit in GCN\\
         130606A & 249.3964 & 29.7963 & YES & 2097  & YES & 14785 & Visible only in I band\\
         130610A & 224.4203 & 28.2072 & YES & 207 & YES &14843 &Analysis in this paper\\
         130612A & 259.7941& 16.7200 & YES &  178 & YES & 14875&Analysis in this paper\\
         GRB:130702:1 & 217.308 &  15.774& - &  - & - & -&No other information \\
         130824 & 288.805 & 10.956 & NO &  2253 & YES& - & Not a GRB\\
         131004A & 296.11 & -2.95 & YES &  95 & YES &  15306&Too faint for RINGO3(R>17) \\
         GRB:576238:0 & - & - & - &  - & - & -\\
         131030A & 345.065 & -5.36 & YES &  3748 & YES &15406& WCS error in RINGO3 observations \\
         140206A &  145.33 & 66.76 & YES & 147 & YES & 15806& Issue with RINGO3 observations \\
         GRB:592204:0  & - & - & - & - &  - & -&Not a GRB\\
         INTEGRAL:GRB:6599:0 & - & - & - & - &  - &-& Not a GRB \\
         GRB:596958:0 & 202.928 &29.258 & - & 206 & YES & -&WCS error \\
         140430A &102.9359& 23.0237 &YES & 123 & YES & 16192 &Analysis in this paper \\
         140516A &252.98 & 39.96 & NO & 3158 & YES & -&- \\
         140709A & 304.66 & 51.22 & YES & 101 & YES &- &Faint for RINGO3 (R>17)\ \\
         141026A & 44.084 & 26.928 & Maybe & 196 & YES &- & Faint for RINGO3 (R>17)\\\
         141220A & 195.0657 & 32.1464 & YES & 128 & YES & 17199& Analysis in this paper \\
         141225A &138.77 & 33.79 & YES & 278 & YES & 17231 &Faint for RINGO3 (R>17)\\
         150302A & 175.53 & 36.811 & NO & 169 & YES & - &-\\
         150309A & 277.10 & 86.42 & NO & 210 & YES & 17556 &-\\
         150317A & 138.98 & 55.46 & Maybe &147& NO & -&No source in the image \\
         150428B & 292.63 & 4.125 &  NO & 172 &  YES & - &-\\
         GRB:650221:0 & 7.256& 59.596 & NO &  310& YES & - &Not a GRB \\
         150831B &271.03 & -27.25 & NO & 183& YES & - & -\\
         150908 & 288.80 & 10.94 & NO & 1273  & YES & - &Not a GRB\\
         151118A & 57.17 & 65.90 & NO & 182 & YES & - &- \\
         151215A & 93.5844 & 35.5159 & YES & 181 & YES & - &Analysis in this paper\\
         160119A & 211.92 & 20.46 & Maybe & 216& YES & - &Faint for RINGO3 (R>17)\\
         160313A & 183.79 & 57.28 & NO & 208& YES & 19177 &- \\
         160316A & 118.92 & -29.56 & NO & 169 & YES & - & Not a GRB \\
         160401A & 89.73 & 26.68 &NO & 168 & YES & 19254 &-\\
         160401B & - & - & NO & 798 & YES & - & Same field as 160401A\\
         160401C &- & - & NO & 3592 & YES &- & Same field as 160401A\\
         GRB:702630:0 & 299.64 & 35.22 & NO &  247& YES &- & Not a GRB\\
         160705B & 168.10 & 46.69 & Maybe & 192 & YES & 19658 & Faint for RINGO3 (R>17) \\
         160714A & 234.49 & 63.80 & NO & 156 & NO & - &-\\
         GRB:704327:0 & 272.61 & 72.05 & NO & 158  & NO & -\\
         160821B & 279.97 & 62.39 & YES & 181 & YES & - & Faint for RINGO3 (R> 17) \\
         161022A & 129.00 & 54.34 & Maybe & 203 & YES & 20090 & Faint for RINGO3 (R> 17) \\
         161214A & 190.72 & 6.83 & YES &  114& YES & 20252 & Faint for RINGO3 (R> 17)\\
         INTEGRAL:GRB:7644:2 & 190.72 & 6.82 & - & 3534 & YES &- & Same as 161214A \\
         170208B & 127.14 & -9.02 & YES &126 & YES &- &Faint for RINGO3 (R> 17)\\
         170604B & 200.80 & 64.19 & NO & 177 & YES & - & -\\
         170728B & 237.98 & 70.12 & YES & 213 & YES & 21375 &Faint for RINGO3 (R> 17)\\
         171003A & 40.91 & 61.43 & NO & 871 & YES & 21961 & Galactic Transient\\
         GRB:778435:0 & 84.08 & 34.44 & NO & 159 & YES & -&Not a new GRB \\
         171020A & 39.24 & 15.20 & YES & 184 & YES & 22033 &Faint for RINGO3 (R> 17)\\
         GRB:782859:0 &40.93 & 61.43 & NO & 934 & YES & - &Not a GRB\\
         171115A & 278.38 & 9.12 & NO & 211 & YES & - & -\\
         180325A & 157.4275& 24.4635 & YES & 146 & YES & 22534 & Analysis in this paper\\
         180512A & 201.93 & 21.40 & NO & 332 & YES &22716 & - \\
         GRB:841583:0 & 245.06 & -15.70 & NO & 1318 & YES & - &Not a GRB\\
         180618A & 169.9410& 73.8371 & YES & 200 & YES &22792 & Analysis in this paper\\
         180704A & 32.66 & 69.96 & NO & 176 & YES & - & - \\
         180720C & 265.63 & -26.62 & NO & 265 & YES & 22991 &- \\
         180904A & 274.24 & 46.62 & NO &  224& YES & 23199 &-\\
         190114C & 54.5048& -26.9464 & YES &201 & YES & - & Analysis in this paper\\
         190427A & 280.21 & 40.30 & NO & 229& YES & - & -\\
         190624A & 144.52 & 46.47 & - & 272 & YES & - &Not a new source\\
         191011A & 44.72 & -27.84 & YES & 140 & NO & - & Faint for RINGO3 (R> 17)\\
         191016A & 30.2695& 24.5099 & YES & 533 & YES & - & Analysis in this paper\\
          \hline
    \end{tabular}}
    
    \label{tab:alert}
\end{table*}

\begin{table*}
    \centering
     \caption{Properties of the GRBs observed by RINGO3 for which we could perform photometry. Bursts from 2014 onwards could also be analysed polarimetrically. We list co-ordinates (J2000) in the second and third column, the fourth column gives MJD start time of our observations, the fifth column gives the time range observed by RINGO3, sixth is the $T_{90}$ for the GRB and the seventh column provides the reference for these numbers, eighth column is the galactic extinction, ninth column is for jet break time ($T_{JB}$), and their references is given in tenth column, redshift is given in eleventh column and their reference is provided in twelfth, and the last column gives the GCN reference of GRB detection.}
     \resizebox{\textwidth}{!}{\begin{tabular}{c|c|c|c|c|c|c|c|c|c|c|c|c}
   
          \hline
           \hline
         GRB & RA ($\degr$) & DEC ($\degr$) & MJD start &$T$-$T_0$ (s) & $T_{90}$ (s) &Ref. &$E(B-V)^{GAL}$ &$T_{JB}$(days) & Ref.& z & Ref. & GCN reference  \\
          \hline
         130606A & 249.3964 & 29.7963& 56449.90 &2097-2697 & 276.6 $\pm$ 19.6 &L16 & 0.021& >1.3 & Y16 &5.91&CT13 &\citet{Ukwatta_2013}\\
         130610A & 224.4203 & 28.2072& 56453.13 &207 -807 & 47.7 $\pm$ 10.7 &L16 &0.0181& >2.9 & E09 &2.0920&S13 &\citet{Cummings_2013}\\
         130612A & 259.7941& 16.7200 & 56455.14& 176-776 & 4.0 $\pm$ 1.4 & L16 &0.065&>1.0 & E09 &2.0060&  T13&\citet{GCN14874}\\
         140430A & 102.9359& 23.0237&56777.85 &124-800&173.6 $\pm$ 3.7&L16 &0.12&>1.15 & K15 &1.6 & K14 &\citet{GCN16190}\\
         141220A & 195.0657 & 32.1464&57011.25& 129-1929 &7.2 $\pm$ 0.47& L16&0.011 &>0.35 & J21 &1.3195 & U14&\citet{GCN17196}\\
         151215A & 93.5844 & 35.5159&57371.12 &182-1982& 17.8 $\pm$ 1.0&G15 & 0.34& >2.3 & E09 &2.59 & X15&\citet{Gibson_2015}\\
         180325A & 157.4275& 24.4635& 58202.07 &147-1947& 94.14 $\pm$ 1.47& T18 & 0.0147& >0.4 & E09 &2.25 & He18 & \citet{troja_2018}\\
         180618A & 169.9410& 73.8371& 58287.02 &200-1400& 3.71$^1$ $\pm$ 0.58 & H18 &0.058&>0.02 & TW & $<$1.2 & S18, J22 &\citet{GCN22790}\\
         190114C & 54.5048& -26.9464& 58497.87 &201-2000 &116.4 $\pm$ 2.56 & H19&0.01& 0.21 & J20 &0.4245 & S19& \citet{GCN23688}\\
         191016A & 30.2695& 24.5099& 58772.17 &3987-7587& 219.70 $\pm$ 183.35&E09 & 0.09& 0.52 & S22,P22 &3.29$\pm$0.4$^2$  & S21 &\citet{GCN26008}\\
          \hline
    \end{tabular}}
    
     {References: L16 - \citet{Lien_2016}; G15 - \citet{Gibson_2015}; T18 - \citet{troja_2018}; H18 - \citet{Hamburg_2018}; H19 - \citet{Hamburg_2019}; E09 - \citet{Evans_2009}; CT13 - \citet{Castro-Tirado_2013}; S13 - \citet{Smette_2013}; T13 - \citet{Tanvir_2013}; K14 - \citet{Kruehler_2014}; 
     U14 - \citet{postigo_2014};X15 - \citet{Xu_2015}; He18 - \citet{Heintz_2018};S18 - \citet{GCN22810}; J22 - Jordana-Mitjans et al. 2022, submitted ;S19 - \citet{Selsing_2019}; S21 - \citet{Smith_2021}; Y16 - \citet{Yasuda_2017}; K15 - \citet{kopac_2015}; J21 - \citet{Jordana_2021}; TW - This Work; J20 - \citet{Jordana_2020}; S22-\citet{Shrestha_2021}; P22-\citet{Pereyra_2022}.\\
     1-This is a short GRB with and extended emission \\
     2-This is photometric redshift all other are spectroscopic redshift.}
    \label{tab:general}
\end{table*}

\section{Data reduction}\label{sec:reduction}

In this section, we present the data reduction technique used to extract counts and uncertainties in counts from eight different images; these values are used to calculate both the photometric and polarization signals. 

\subsection{Photometry and calibration}\label{subsec:reduction-Photometry}

First we perform photometric reduction on the images and extract counts and uncertainties for the eight different images. We perform aperture photometry using the Python package Astropy Photutils \citep{Bradley_2019}. We first detect sources in the field of view (FOV) with a minimum of 15 times the standard deviation of the image signal-to-noise ratio (SNR) using DAOStarFinder. Once we identify these sources, we estimate background noise using Background2D function of Photutils and subtract this from the data. After the background is subtracted, we perform aperture photometry, which requires the selection of the appropriate aperture size. We use two different methods to calculate the best aperture size; 1) calculate full width at half-maximum (FWHM) of the source, and 2) calculate counts and error in counts with respect to different aperture size for one stacked image per observation. We use 2 to 3 times FWHM of the target and the aperture that produces the best counts to counts error ratio as our aperture size to perform photometry per target. We obtain eight different counts and error in counts. Error in counts are calculated via root mean square sum of background noise and Poisson noise of the source \citep{Bradley_2019}. 

We perform relative photometry with USNO-B1.0 catalogue stars to calculate the magnitude of the GRBs. The sum of eight polarized images in a RINGO3 observations provides the total intensity of the source. For each GRB, we get simultaneous observations for three different wavebands. In the same FOV, we select one or two stars whose magnitude is already known and use those sources to calibrate the magnitude of the GRB being observed. Colour transforms from \citet{kopac_2015,Jordana_2020} were used to convert the RINGO3 magnitudes to the standard Johnson-Cousins system. In order to correct for Galactic extinction we used \citet{Schlafly_2011} to correct the magnitude of the GRBs. To convert magnitudes to fluxes we used zeropoint values from \citet{Bessell_1998}.

\begin{figure*}
     \includegraphics[width=2\columnwidth]{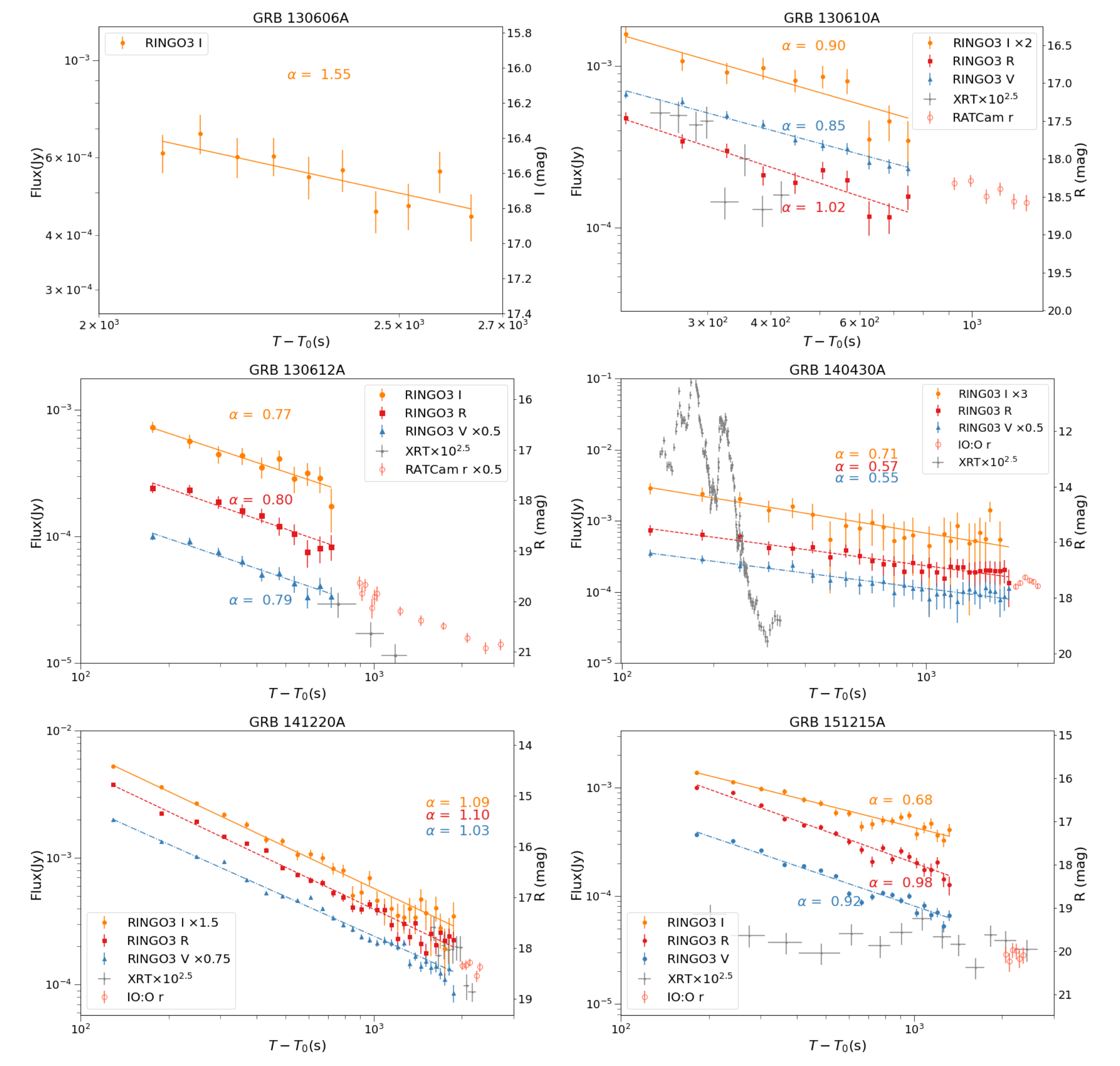}
 
    \caption{Light curves of the first six GRBs observed by RINGO3. Power-law and broken power-law fit are performed on these GRBs according to the data observed. Power-law decay index ($\alpha$) is presented in the plots for three different bands. RATCam, IO:O, and Swift XRT data are plotted as well. Left y-axis and x-axis are in log scale. For all the cases, background is estimated using 2D background estimate.}
    \label{fig:photometry_1}
\end{figure*}

\begin{figure*}
     \includegraphics[width=2\columnwidth]{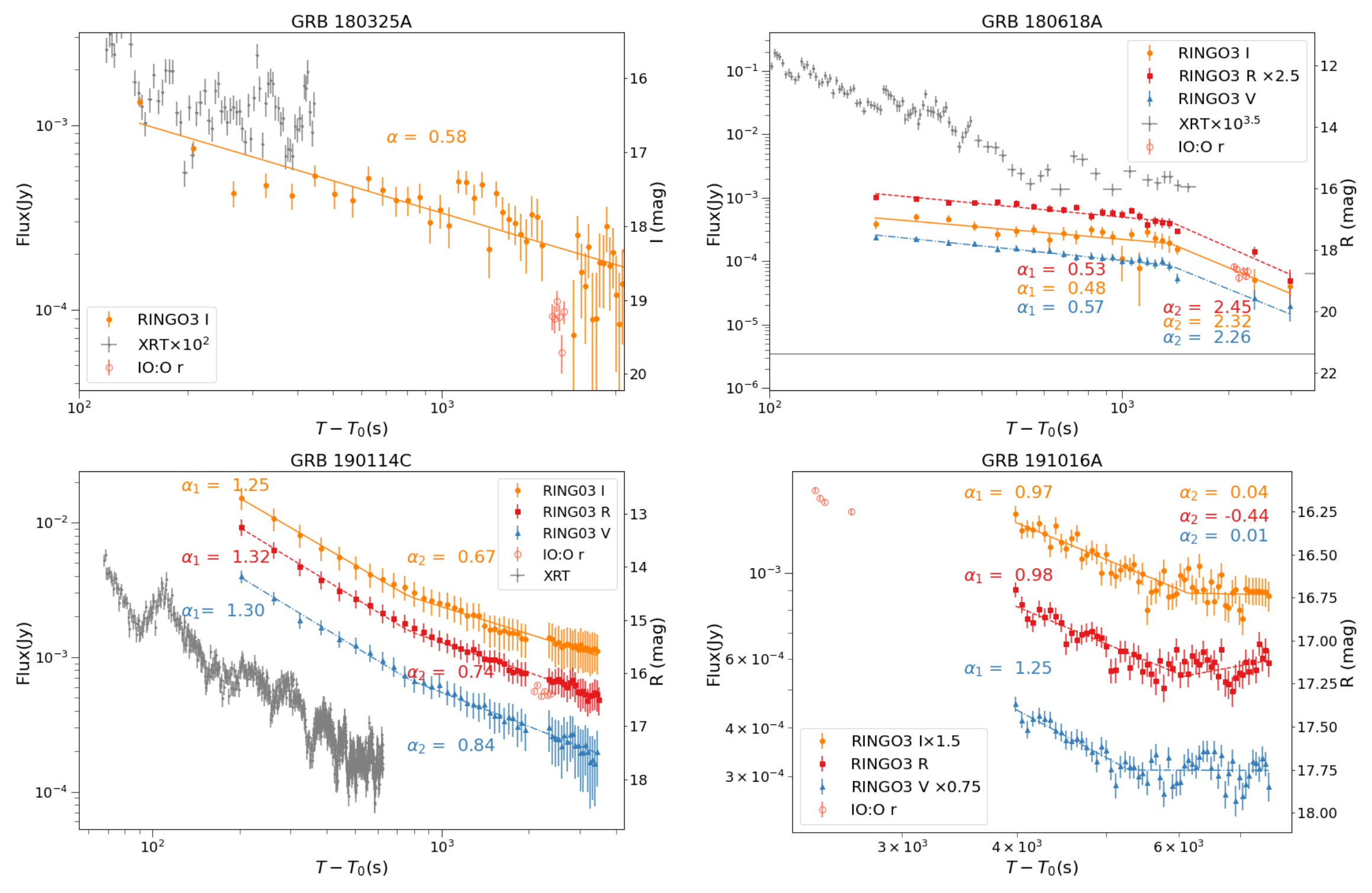}
 
    \caption{Light curves of the last four GRBs observed by RINGO3. Power-law and broken power-law fit are performed on these GRBs according to the data observed. Power-law decay index ($\alpha$) is presented in the plots for three different bands. RATCam, IO:O, and Swift XRT data are plotted as well. Left y-axis and x-axis are in log scale. For all the cases, background is estimated using 2D background estimate.}
    \label{fig:photometry_2}
\end{figure*}

  Properties such as redshift and high energy burst duration from literature for all ten GRBs are presented in Table~\ref{tab:general}. In Fig.~\ref{fig:time_coverage} we present the time coverage of GRB observations by RINGO3 in observer frame in red and in co-moving frame in grey. The BAT peak time and $T_{90}$ in the observer frame are presented as triangles and circles in the same plot for all ten GRBs. The light-curves of all ten GRBs observed are presented in Fig.~\ref{fig:photometry_1} and Fig.~\ref{fig:photometry_2}. We fit the optical light curves of RINGO3 data with either a simple (PL) or broken power (BPL) and provide reduced $\chi^2$ value for each fit in the Tables~\ref{tab:photometry_pl} and ~\ref{tab:photometry_bpl} respectively. We perform a PL or BPL fit for each wave band observation separately, thus giving us different decay indices ($\alpha$) for different wavelength observations. For GRB 191016A, we perform a PL fit for IO:O data as well to see the earlier time decay index compared to later time observations made by RINGO3.

\begin{table*}
    \centering
    \begin{tabular}{c|c|c|c|c|c|c|c|c}
          \hline
           \hline
         GRB & Model & $\alpha I $ & $\alpha R $  & $\alpha V$ & $\chi_{r}^2 (I)$ &$\chi_{r}^2 (R) $ &$\chi_{r}^2 (V)$ & d.o.f\\
         
          \hline
         130606A  & PL & 1.55 $\pm$0.5 & - & - & 6.1  & - & - & 8\\
         130610A & PL & 0.90$\pm$0.13 & 1.02 $\pm$ 0.1 & 0.85 $\pm$ 0.05& 2.2  & 0.85  & 2.7   & 8\\
         130612A & PL & 0.77 $\pm$0.07 & 0.80 $\pm$ 0.07 & 0.79 $\pm$ 0.06& 1.2 & 1.4 & 1.3 & 8 \\
         140430A & PL & 0.71 $\pm$ 0.06& 0.57 $\pm$ 0.02 & 0.55 $\pm$ 0.02 & 0.4 & 0.2 & 0.2 &28\\
         141220A & PL & 1.09 $\pm$ 0.02&  1.10 $\pm$ 0.02& 1.03 $\pm$ 0.02& 1.05 & 2.4 & 4.3  &28 \\
         151215A & PL & 0.68 $\pm$ 0.03 & 0.98 $\pm$ 0.03 & 0.92 $\pm$ 0.03 & 1.7 & 2.9 & 6.0  & 16 \\
         180325A  & PL & 0.58 $\pm$ 0.04 & -& -& 2.04 &- & - &48 \\
         
          \hline
    \end{tabular}
    \caption{Light curve fitting results for GRBs that can be fitted with single power-law (PL) model. PL is given by $F \propto t^{-\alpha}$ For each fit we provide reduced $\chi_{r}^2$ values and degree of freedom (d.o.f). }
    \label{tab:photometry_pl}
\end{table*}

\begin{table*}
    \centering
    \resizebox{\textwidth}{!}{\begin{tabular}{c|c|c|c|c|c|c|c|c|c|c|c|c}
          \hline
           \hline
         GRB & Model & $T_{b}$ (m) &$\alpha_1 I $ & $\alpha_2 I $ & $\alpha_1 R $  & $\alpha_2 R $  & $\alpha_1 V$ & $\alpha_2 V$ & $\chi_{r}^2 (I)$ &$\chi_{r}^2 (R) $ &$\chi_{r}^2 (V)$ & d.o.f\\
         
          \hline
180618A  & BPL & 22.8 &0.48$\pm$ 0.08 & 2.32 $\pm$ 0.8  & 0.53$\pm$ 0.04  & 2.45$\pm$ 0.4 & 0.57$\pm$ 0.05 & 2.26$\pm$ 0.5& 1.2 & 1.4 &0.6 &18 \\
 190114C  & BPL & 6.7 &1.43$\pm$ 0.03 & 0.87 $\pm$ 0.02 & 1.50$\pm$ 0.02 & 0.94 $\pm$ 0.01 & 1.47$\pm$ 0.02 & 0.99 $\pm$ 0.02 & 0.46 & 0.3 & 0.08 & 45\\
 191016A  & BPL & 102.4 ($I$), 101.4 ($R$), 87.5 ($V$) &0.97$\pm$ 0.07 & 0.04$\pm$ 0.17   & 0.98$\pm$ 0.07  & -0.44$\pm$ 0.17 & 1.25$\pm$ 0.1  & 0.01$\pm$ 0.09 & 0.9 & 1.01 & 1.04 & 56 \\
 
 \hline
    \end{tabular}}
    \caption{Light curve fitting results for GRBs whose light-curve is fitted by broken power-law (BPL) model. $\alpha_1$ and $\alpha_2$ denotes the decay index before and after the break time respectively. For each fit we provide reduced $\chi_{r}^2$ values and degree of freedom (d.o.f). }
    \label{tab:photometry_bpl}
\end{table*}

\subsection{Polarization signal}\label{subsec:reduction-polarimetry}
The sky-subtracted counts of the eight different images are used to extract polarization of the source using recipe shown by \citet{clarke_2002}. The same process is followed for RINGO, RINGO2, and previous RINGO3 analysis (e.g. \citet{Jermak_2016, Shrestha_2021}). Using this technique, we can get linear Stokes parameters $q$ and $u$.

In every case we need to correct for polarization introduced by the instrument itself, therefore we observe different unpolarized and polarized standards. We take the average of Stokes q and u for the unpolarized standard star, with the assumption that the unpolarized standard star has Stokes $q \sim 0$ and $u \sim 0$, and calculate the average values introduced by the instrument. With the instrumental Stokes parameters being $q_{inst}$ and $u_{inst}$, the instrument corrected Stokes parameters of the target are given by:
\begin{align}
    & q_c = q - q_{inst} \\
    & u_c = u - u_{inst}.
\end{align}

We perform error propagation in $q$ and $u$ to calculate the error value $q_e$ and $u_e$. Finally, these $q_c$ and $u_c$ are used to calculate a raw percentage polarization and position angle by using:

\begin{align}
    & \%p = \sqrt{q_c^2+u_c^2} \times 100. \label{eq:pol} \\
    & \psi =\frac{1}{2} \arctan\left ( \frac{u}{q} \right).
    \label{eq:pa}
\end{align}

Analysis of RINGO3 data of polarized standard shows no significant instrumental depolarization \citep{Jermak_2017}.
 
We take measures to get the correct quadrant for the position angle. The position angle needs to be rotated based on the telescope Cassegrain axis sky position angle (SKYPA), measured east of north which gives electron vector position angle  (EVPA).  
\begin{equation}
    EVPA = \psi + SKYPA + K.
    \label{eq:evpa}
\end{equation}

Here $K$ is a calibration factor which gives the position angle offset combined of the angles between the orientation of the polarizer, the telescope focal plane, and the trigger position of the angle measuring sensor. This angle offset was calculated using polarized standards observed during various time periods and the values of $K$ are provided in Table~\ref{tab:inst}. Some parts of the analysis are taken from \citet{Jermak_2017}.

The last step in the polarization calculation is bias correction and error calculation for the polarization degree and EVPA. Noise in \textit{q} and \textit{u} introduces a polarization signal which is not intrinsic. In order to correct for this and to calculate the error in polarization degree, we use the prescription developed by \cite{Plaszczynski_2014}. 
The error in EVPA is calculated using standard error propagation applied to Eq.~\ref{eq:pa}. 

\begin{table*}
    \centering
    \begin{tabular}{c|c|c|c|c|c|c}
          \hline
           \hline
         MJD Range  & $q_{in}$ (I)& $\sigma$  & $u_{in}$(I) & $\sigma$ & K (I $\degr$) & K (I $\degr$ $\sigma$) \\
          \hline
         56658-56816& -0.0119$\pm$0.0005 &0.003 & -0.0410 $\pm$ 0.0006 & 0.0036& 57.39 & 4.26 \\
         56816-57202  & -0.0154 $\pm$ 0.0004 & 0.004 & 0.0295$\pm$ 0.0015& 0.014 & 115.15 & 3.75\\
         >57202 & -0.0131 $\pm$ 0.004& 0.025  &  -0.0336 $\pm$ 0.001 &0.006   & 125.61 & 4.63 \\
          \hline
           MJD Range  & $q_{in}$ (R)& $\sigma$  & $u_{in}$(R) & $\sigma$ & K (R $\degr$) & K (R $\degr$ $\sigma$) \\
          \hline
         56658-56816&  -0.01163  $\pm$ 0.0004& 0.0024 &  -0.0371 $\pm$ 0.0048& 0.029 &  55.58 & 3.6 \\
         56816-57202  & -0.0157 $\pm$ 0.0003 & 0.002 & 0.0333 $\pm$ 0.0014 & 0.013 &  115.95 & 3.25\\
         >57202 & -0.0105 $\pm$ 0.0024&0.014   &  -0.0356 $\pm$ 0.009& 0.059  & 124.8 & 5.05 \\
         \hline
         MJD Range  & $q_{in}$ (V)& $\sigma$  & $u_{in}$(V) & $\sigma$ & K (V $\degr$) & K (V$\degr$ $\sigma$) \\
          \hline
         56658-56816& -0.0096 $\pm$ 0.0003& 0.002 &-0.0177 $\pm$ 0.0003&  0.002& 54.93 & 3.94 \\
         56816-57202  & -0.0077 $\pm$ 0.0002& 0.0018  &  0.0215 $\pm$ 0.0009 & 0.008 &  115.42 & 2.92\\
         >57202 & -0.0047 $\pm$ 0.004& 0.026  &  -0.022 $\pm$ 0.0058& 0.035  & 124.90 & 4.89 \\
         \hline
         
    \end{tabular}
    \caption{Table of instrument $q$, $u$, and K factor values for different time periods of observations using RINGO3. We quote standard error ($\frac{\sigma}{\sqrt{N}}$) where $N$ is the total number of observations as the error in instrument $q$ and $u$. Standard deviation ($\sigma$) is also presented. These values are much smaller than error in Stokes $q$ and $u$ of GRB.  Instrumental $q$, $u$, and $K$ values are from \protect\cite{Jermak_2017}.  }
    \label{tab:inst}
\end{table*}

\begin{table*}
    \centering
    \begin{tabular}{c|c|c|c|c|c|c|c|c|c|c|c|c}
     \hline
      \hline
         GRB & $T$-$T_0$ (s) & P (\%) ($I$) & P (\%) ($R$) & P (\%) ($V$) & Rank ($I$) & Rank ($R$) & Rank ($V$)\\ 
          \hline
         140430A &124-724& <35.9&<18.9 & <18.50.97 & 0.67 & 0.20 \\
         140430A &724-1324& <42.2&<58.6 & <32.3& 0.78 & 0.68 & 0.66 &\\
         140430A &1324-1924 & <96.91&<24.1 &<23.0& 0.78 & 0.54 & 0.30 \\
         \hline
         140430A &124-1924 & <12.6 &<9.6 &<16.5& 0.97 & 0.11 & 0.099 \\
         \hline
         \hline
         141220A & 129-729 & <17.9& <10.8& <7.5& 0.88 & 0.67 & 0.79  \\
         141220A & 729-1329 & <15.3& <6.7& <11.1& 0.45 & 0.69 & 0.21  \\
         141220A & 1329-1929 & <69.5& <31.7& <11.2 & 0.84& 0.76 & 0.57\\
         \hline
         141220A & 129-1929 & <3.52& <3.23& <1.74 & 0.33 & 0.13 & 0.73 \\
         \hline
         \hline
         151215A & 182-782& <7.8& <5.9 & <3.3 & 0.31 & 0.199 & 0.065 \\
         151215A & 782-1382& <14.0& <20.27 & <4.2& 0.16 & 0.73 & 0.051  \\
         151215A & 1382-1982& <22.34&<28.9 & <5.6&   0.67 & 0.37 & 0.95 \\
         151215A & 1982-2582& <15.2& <22.5 & <8.2 & 0.047 & 0.044 & 0.32 \\
         \hline
         151215A & 182-2582& <6.45& <6.64 & <6.05 & 0.059 & 0.91&0.91 \\
         \hline
         \hline
         180325A & 147-747&<18.2 &-& -& 0.70 &-&-\\
         180325A & 747-1347&<6.8 &-&-&  0.20 &-&-\\
         180325A & 1347-1947&<30.1 &-&-&  0.24 
         &-&-\\
         \hline
          180325A & 147-1947&<12.29 &-&-&  0.342 &-&- \\
         \hline
         \hline
         180618A &800-1400& <66.0& <25.7 &<10.7  & 0.90 & 0.95 & 0.54 &\\
         180618A &1400-2000& <71.4&<19.5 &<25.5  & 0.95 & 0.55 & 0.107 &\\
         180618A &2000-2600&<100&<100 &<26.0 & 0.88 & 0.78 & 0.53 \\
         \hline
         180618A &800-2600& <5.26&<12.24 &<6.78  & 0.38 &0.61 & 0.078 \\
         
         \hline
         \hline
         
          \hline
          \hline
    \end{tabular}
    \caption{Table of polarization degree measurements from the observation done by RINGO3. The upper limit of polarization for each time interval is presented along with the upper limit for stacked data of the whole observation is presented for each GRB. The presented data are not Milky Way ISP corrected. Columns are GRB identifier, time range, polarization degree for $I$, $R$, and $V$ bands,  permutation rank for detected polarization.}
    \label{tab:polarization_nondetection}
\end{table*}

\begin{table*}
    \centering
    \begin{tabular}{c|c|c|c|c|c|c|c|c|c|c}
     \hline
      \hline
         GRB & $T$-$T_0$ (s) & P (\%) ($I$) & P (\%) ($R$) & P (\%) ($V$) & EVPA (\degr) ($I$) &  EVPA (\degr) ($R$) &EVPA (\degr) ($V$) & Rank ($I$) & Rank ($R$) & Rank ($V$)\\ 
          \hline
         \hline
         190114C &203-803& {$\bf 2.9 \pm 0.8$}&{$\bf 3.2 \pm 0.8$} &{$\bf 2.0 \pm 1.2 $} & 26 $\pm$ 9& 48 $\pm$ 9 & 25 $\pm$ 25 & 0.98 &0.997& 0.99 \\
         190114C &803-1403& 2.0 $\pm$ 1.5 & 2.5 $\pm$ 2 &<3.7 & - & -&-&  0.147 &0.32 & 0.87 \\
         190114C &1403-2003& 3.7 $\pm$ 2.6&<4.8 &<5.0 & - &-&- & 0.32 & 0.07 & 0.801 \\
         190114C &2354-2954& <4.1&<3.1 &<6.5 &- &-&-& 0.156 & 0.82 & 0.99 \\
         190114C &2954-3554& <8.2&<10.5 &<6.4 &- &-&-& 0.87 & 0.81 & 0.98 \\
         \hline
         190114C &203-2003& <2.7&<2.8 &<2.22 &- &-&-& 0.82 & 0.79 & 0.53 \\
         
         \hline 
         \hline
         191016A & 3987-4587& $\bf 4.7 \pm 4.1$ & $<9.1$ &  $\bf 7.8 \pm 5.6$ & 93 $\pm$ 22 &-& 90 $\pm$ 18 & 0.99 & 0.63 & 0.99\\
         191016A & 4587-5187&$<5.2$ &$\bf 11.2 \pm 6.6$ & $\bf 5.7 \pm 5.6$&-& 90 $\pm$ 15 & 82$\pm$ 26  & 0.59 & 0.99 &0.98 \\
         191016A & 5187-5787&<14.0 & $<5.5$& <10.8&-&-&- & 0.72 & 0.11 & 0.73  \\
         191016A & 5787-6387& $\bf 14.6 \pm 7.2$ & $\bf 6.1 \pm 6.1$ & <9.2 & 100 $\pm$ 12 & 90 $\pm$ 30 &- &0.99 & 0.95 & 0.74 \\
         191016A & 6387-6987& <10.7 & <12.0& <13.5&-&-&-& 0.67 & 0.83 & 0.86 \\
         
         191016A & 6987-7587&<17.6 & <11.0& <9.2 & - &-&- & 0.76 & 0.86 & 0.94 \\
         
         \hline
          191016A & 3987-7587&<3.82 & <5.23&<3.7&-&-&-  & 0.34 & 0.141 &  0.097\\
          \hline
          \hline
    \end{tabular}
    \caption{ Table for two GRBs with polarization detection. The polarization degree for each time interval is presented and the upper limit for stacked data for the whole observation time range is also presented. The values presented are not corrected for Milky Way ISP contribution. Columns are GRB identifier, time range, polarization degree for $I$, $R$, and $V$ bands, Position angle, permutation rank for detected polarization. Detections that pass both error bar and permutation analysis are highlighted in bold. }
    \label{tab:polarization_detection}
\end{table*}

\begin{table*}
    
    \centering
    \begin{tabular}{c|c|c|c|c|}
          \hline
           \hline
         GRB & $E(B-V)^{GAL}$ & $I-ISP^{GAL} (\%)$ & $R-ISP^{GAL} (\%)$ & $V-ISP^{GAL} (\%)$   \\
          \hline
         140430A & 0.14 & 1.03& 1.17&1.26\\
         141220A &  0.01&0.07 &0.08 &0.09\\
         151215A &  0.40&2.95&3.35 & 3.6\\
         180325A &  0.02&0.14 &0.16 &0.18\\
         180618A & 0.07&0.51 &0.58 &0.63\\
         190114C & 0.01& 0.07& 0.08 & 0.09\\
         191016A &  0.09&0.66 & 0.75 & 0.8\\
          \hline
    \end{tabular}
 \caption{Table of upper limit of Galactic interstellar polarization estimates based on extinction values from \citet{Schlafly_2011}} and using\ \protect\cite{serkowski_1975}.
    \label{tab:isp}
\end{table*}

\begin{figure*}
\centering
\includegraphics[width=2\columnwidth]{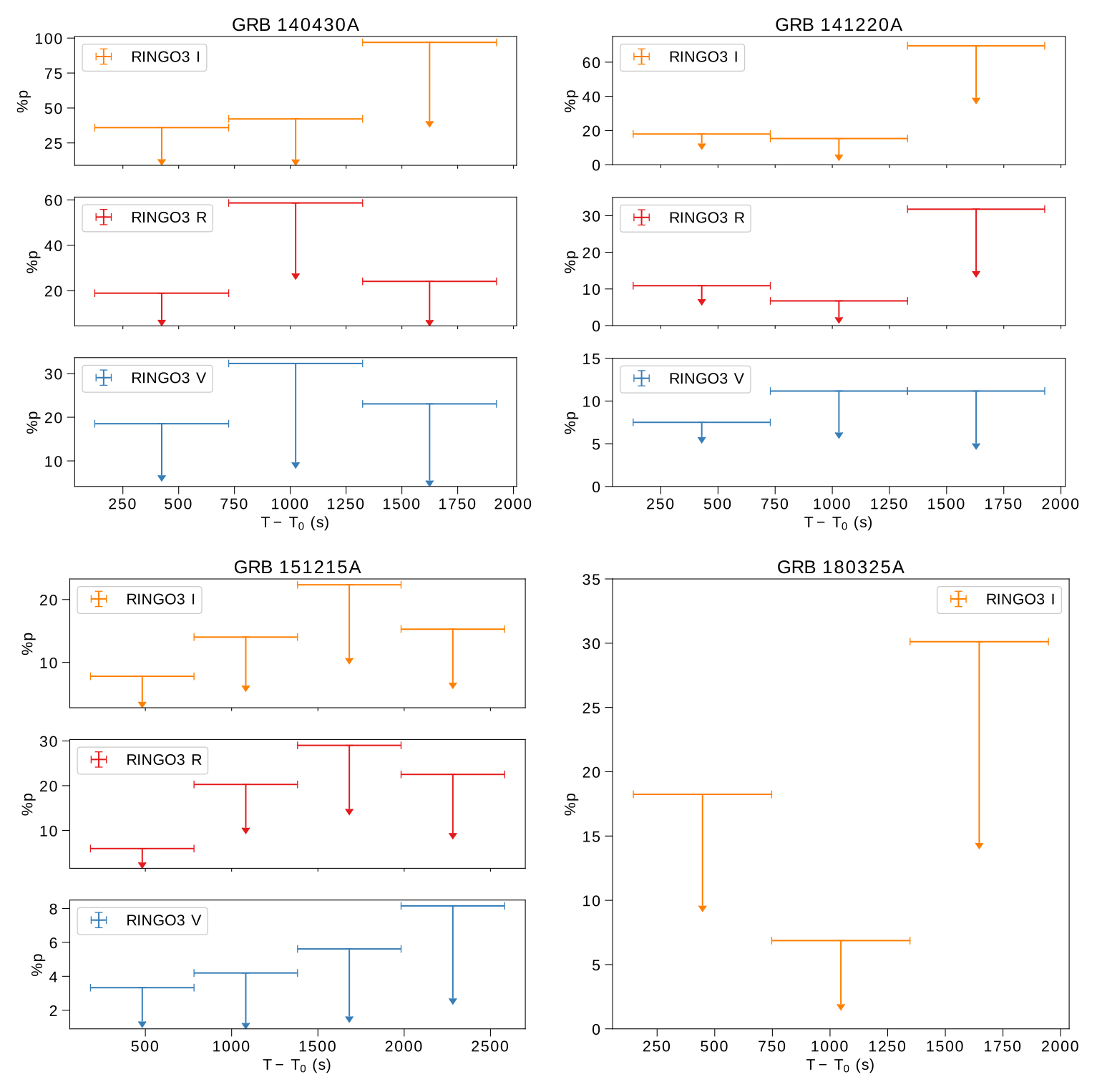}
\caption{Observed polarization degree with respect to $T-T_0$ for the first four GRBs. For each GRB, the results in $I$, $R$, and $V$ bands are presented in the top, middle, and bottom panels respectively. Upper limits and detections are presented accordingly. Uncertainties on the x-axis are the binned exposure times for the given data point.  The results presented here are not corrected for interstellar polarization. }
\label{fig:pol_results_1}
\end{figure*}

\begin{figure*}
\centering
\includegraphics[width=2\columnwidth]{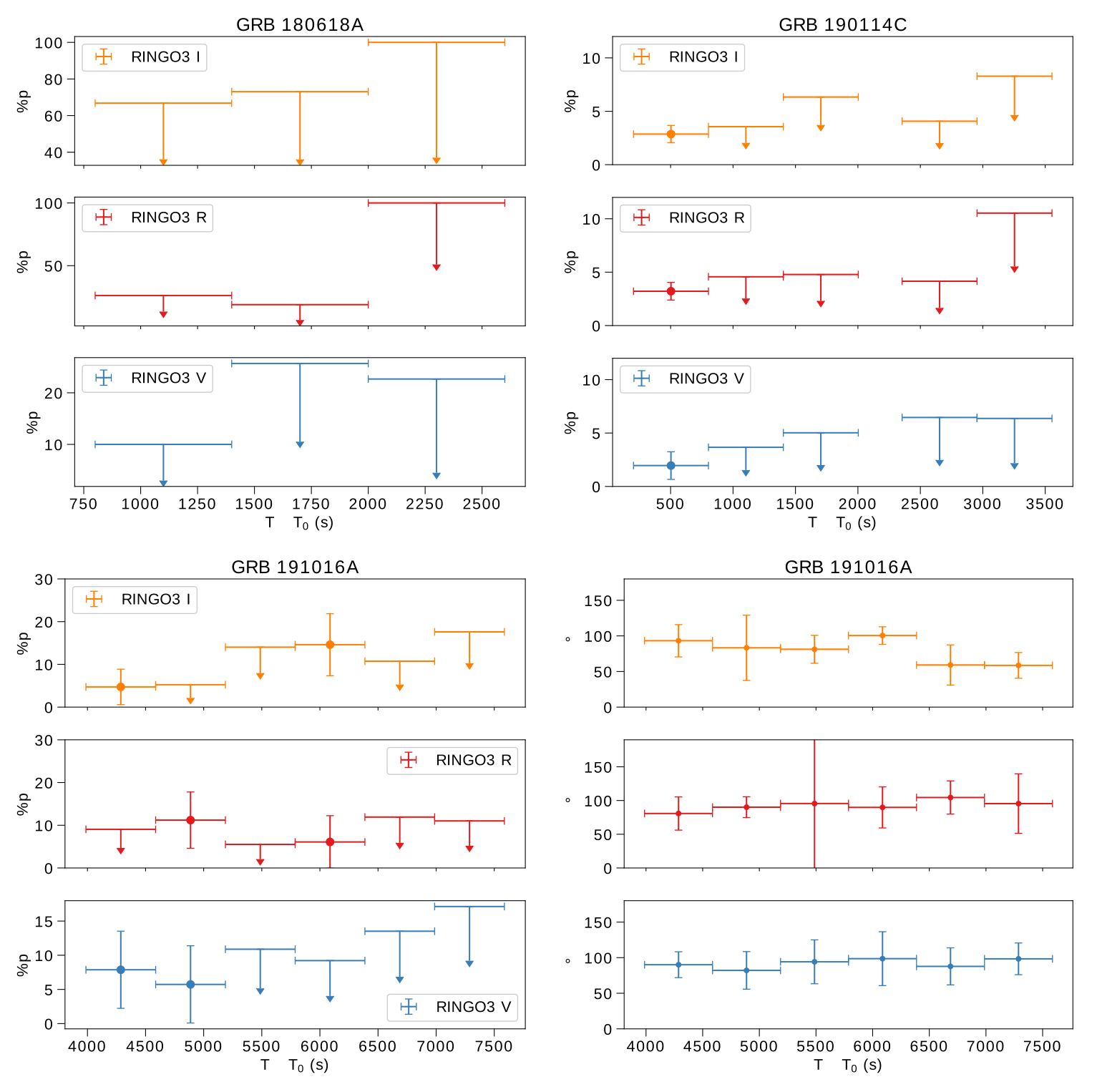}
\caption{Observed polarization degree with respect to $T-T_0$ for the last three GRBs. For each GRB, the results in $I$, $R$, and $V$ bands are presented in the top, middle, and bottom panels respectively. Upper limits and detections are presented accordingly. Uncertainties on the x-axis are the binned exposure times for the given data point. Only for GRB 191016A, the position angle plot is presented. The results presented here are not corrected for interstellar polarization.  }
\label{fig:pol_results_2}
\end{figure*}

In Figures~\ref{fig:pol_results_1} and ~\ref{fig:pol_results_2} , we present the evolution of polarization with time for all seven GRBs for the three different wavebands $I$, $R$ and $V$ (top to bottom panels). We present $1\sigma$ error bars in these plots for the polarization results. When the error bars in polarization degree crosses $0 \%$, then we consider the polarization to be an upper limit and upper limit value is presented by the upper end of the error bar. When we have error bars not crossing $0\%$, we consider it to be a possible detection.

\subsection{Polarization Detections}

Table \ref{tab:polarization_detection} identifies possible polarization detections in GRB 190114C at early times (before 2003-s) and in GRB 191016A at various epochs as from consideration of their error bars as outlined above.

In order to confirm the possible polarization detections, we implemented ``permutation analysis" (described in detail in \cite{steele-ringo2-2017}) to rigorously investigate the probabilities of the detection by looking at the individual counts at the 8 rotor positions for a source.  Before doing this we correct for instrumental polarization using a bright star in the field of view as an unpolarized source and dividing the GRB source counts by the unpolarized source counts at the corresponding rotor position.  These corrected counts from the eight rotor positions are shuffled into all possible ordered permutations.  This procedure will destroy any coherent polarization signal encoded in the data and generates $(8-1)!$ (5040) permutations of the corrected counts for the GRB source.  Each of these permutations will have identical noise characteristics to the original data (being generated directly from it) and are then used to calculate a polarization degree.  By sorting the resulting polarization values we can then generate a rank which tells us the probability of the detected polarization degree being true (as opposed to being artificially created by the transformation of noise into polarization signal due to polarization bias). We can then check the null hypothesis; if the source is unpolarized then what is the chance of getting some polarization signal due to noise in the data? For example if the rank is greater than 0.9 it means the probability of being an unpolarized source $p = 1-rank$ will be $<0.1$.

Since we have carried out a total of 79 tests over the sample, using the threshold $p<0.05$, we could of course expect $\sim4$ false positives to have arisen from this testing procedure under the null hypothesis that all GRBs do not show polarization. Overall we find a total of 13 such positives in the sample. The binomial cumulative probability of such an outcome is highly significant ($p<1.e\times10^{-4}$) indicating that at least some of our detections ($\sim 9$) should be true polarization signals.

Further evidence of the validity of the detections can be inferred from the correspondence between individual detections made by the two techniques (error bar analysis and permutation analysis).  From the error bar analysis we find 13 measurements are identified as possible detections, 9 of which have permutation analysis with $p<0.05$.  Testing against the null hypothesis of no correspondence, we find the cumulative binomial probability of this outcome is $p<1\times10^{-6}$, indicating a strongly significant association.   In comparison only 4 out of the 65 measurements which have only error bar upper limits show a permutation $p<0.05$ - an outcome with a cumulative binomial $p<0.41$ i.e. entirely consistent with the high permutation rank values in this case being spurious due to the multi-trial nature of the test. Overall we are therefore confident that the 9 measurements that pass both techniques (error bars and permutation analysis) are true detections of polarization.  We highlight these measurements in bold text in Table \ref{tab:polarization_detection}.

\subsection{Galactic Interstellar Polarization estimate}
We use known GRB Milky Way Galactic extinction values to estimate the Milky way Galactic interstellar polarization (GISP) following the formulation by \citet{serkowski_1975}. First we used $p^V(GISP) \leq 9 E_{B-V}$ \footnote{https://irsa.ipac.caltech.edu/cgi-bin/bgTools/nph-bgExec} to calculate the upper limit in polarization induced by GISP in the $V$ band. For each GRB in our observed sample we used $5\degr \times 5\degr$ statistics from \citet{Schlafly_2011}. After this calculation, we used $p/p_{max} =\exp[-\kappa {\rm ln}^2(\lambda_{max}/\lambda)]$ to calculate GISP in $R$ and $I$ bands using $V$ band as the $\lambda_{max}$ and $p^V$ as the $p_{max}$; where $p$ is the polarization induced by GISP at the wavelength $\lambda$, $p_{max}$ is the maximum polarization induced by GISP at the wavelength $\lambda_{max}$, and $\kappa$ (normally $K$ is used but here to avoid confusion with $K$ of EVPA constant we use $\kappa$)is a constant given to be 1.15 in \citet{serkowski_1975} and \citet{Wilking_1982} later modified it to be $\kappa = -0.10+1.86 \lambda_{max}$, this is used for our analysis in this paper. The GISP estimates for the seven GRBs are presented in Table~\ref{tab:isp}. We do not correct for GISP in our polarization results as the GISP values are very low and can only be used for reference. We also do not correct for the ISP contribution from the host galaxy for all the GRBS. However, we present the host galactic ISP contribution for GRB 191016A and GRB 190114C, for which we have a probable detection.

\section{Results}\label{sec:results}
In this section we present the photometric results of ten GRBs and polarimetric results of seven GRBs. In Table~\ref{tab:flux}, we provide a subset of flux values for GRB 130606A and full data for all the GRBs are available in machine readable format online.

\begin{table}
\begin{tabular}{llrrrr}
\hline
    GRB & Filter &  $T_{start}$(s) &  Exp(s) &  $F_\nu$ (mJy) &  $F_\nu{err}$(mJy) \\
\hline
130606A &      I &        2097.0 &      60 &       0.615 &              0.061 \\
 130606A &      I &        2157.0 &      60 &       0.681 &              0.070 \\
 130606A &      I &        2217.0 &      60 &       0.603 &              0.063 \\
 130606A &      I &        2277.0 &      60 &       0.605 &              0.061 \\
 130606A &      I &        2337.0 &      60 &       0.543 &              0.060 \\
 130606A &      I &        2397.0 &      60 &       0.563 &              0.062 \\
 130606A &      I &        2457.0 &      60 &       0.453 &              0.050 \\
 130606A &      I &        2517.0 &      60 &       0.467 &              0.056 \\
 130606A &      I &        2577.0 &      60 &       0.558 &              0.060 \\
 130606A &      I &        2637.0 &      60 &       0.441 &              0.054 \\
\hline
\end{tabular}

\caption{Sample photometry of GRB 130606A. Here columns are GRB name, RINGO3 Filter, $T_{start}$ is the start time of exposure in seconds since the trigger time, $T_{exp}$ is exposure time in seconds, $F_\nu$ is flux in mJy, and $F_\nu err$ is error in flux in mJy. This table is available in its entirety in machine-readable form.}
    \label{tab:flux}
\end{table}

\subsection{GRB 130606A}

RINGO3 observations of this burst were obtained $\sim$35 minutes after the trigger time of 21:04:39 UTC \citep{Ukwatta_2013} and only the $I$ band has SNR high enough to perform photometry. The gamma-ray duration is $T_{90} = 276 s \pm 20$ \citep{Lien_2016} and spectroscopic redshift of $z = 5.91$ as observed by GTC \citep{Castro-Tirado_2013}. We corrected for Galactic extinctions for the GRB corresponding to  $A_V =0.064 $, $A_R =0.051$, and $A_I = 0.036 $. We fit a power-law to the light curve and get a decay index of 1.55 $\pm$ 0.5 as shown in the Fig.~\ref{fig:photometry_1} and Table~\ref{tab:photometry_pl}. Swift XRT data are best fitted by 2 breaks at $482^{+243}_{-393}$ s and $2.1 ^{+0.6}_{-0.3} \times 10^4 $ s with a decay index of $0.63^{+0.09}_{-1.77}$, $1.09 \pm 0.05$, and $1.79^{+0.22}_{-0.19}$ \citep{Evans_2009}. \citet{Yasuda_2017} constraint the jet break time to be greater than 1.3 days. Polarimetric analysis are not presented here because we could not constrain the instrument polarization during this period.

\subsection{GRB 130610A}
RINGO3 photometric observations were obtained in $I$, $R$, and $V$ bands $\sim$3 minutes after the trigger time of 3:12:13 UTC and RATCam observations were obtained in SDSS $g'$,$r'$, and $i'$ band $\sim$18 minutes after the trigger. We used stars in the field to do photometric calibrations and corrected for Galactic extinction corresponding to $A_V =0.058 $, $A_R =0.046$, $A_I = 0.033 $, $A_{g'} = 0.071  $, $A_{r'} =0.049 $, and $A_{i'} =0.037 $. A simple power-law is fitted to all the data in three different wave bands. We get $\alpha = 0.90 \pm 0.13, 1.02 \pm 0.1,$ and $0.85 \pm 0.05$ for $I$, $R$ and $V$ filters. We note that for R band images there is a dark line running through the image where the GRB is located which has been seen previously in RINGO3 R images. We found that the light-curve behaviour is dependent on how we subtract the background noise. When the median of the image is considered the background, we get the decay index to be $0.56 \pm 0.05$ which is much lower than the value we get when we use a 2D background estimate.We present results in this paper using 2D background estimate. Swift XRT and RATCam data are also presented in the Fig.~\ref{fig:photometry_1} along with RINGO3 data and the power-law fit.
This GRB has $T_{90} =47 \pm 11 s$ \citep{Lien_2016} and a spectroscopic redshift of $z = 2.092$ \citep{Smette_2013}. The Swift XRT light-curve was fitted by one break at $242^{+27}_{-33}$ seconds with a decay index of $2.47^{+0.44}_{-0.29}$ and $1.09 \pm 0.03$ before and after the break \citep{Evans_2009}. From the Swift XRT light-curve, we assume the minimum jet break time to be 2.9 days \citep{Evans_2009}. 

\subsection{GRB 130612A}
This is the one of the GRBs in the sample that might be argued to be a short burst ($T_{90} = 4.0  \pm 1$ s \citep{Lien_2016}) and was observed around 3 minutes after the trigger time of 3:22:23.361 UTC as reported by Swift. RINGO3 photometric observations were made in $I$, $R$, and $V$ bands and RATCam observations were obtained in SDSS $g'$, $r'$, and $i'$ band. Calibration was done using stars in the field and the Galactic extinction correction corresponded to $A_V =0.204 $, $A_R =0.161$, $A_I = 0.115$, $A_{g'} = 0.251 $, $A_{r'} =0.174 $, and $A_{i'} =0.129 $. Fig.~\ref{fig:photometry_1} shows the light curve of GRB 130612A for all three filters of RINGO3 including RATCam and Swift XRT data. Power-law fits are applied to RINGO3 data with $\alpha$ = $0.77 \pm 0.09$, $0.85 \pm 0.09$, and $0.80 \pm 0.06$ for $I$, $R$ and $V$ filters respectively. Swift XRT data was best fitted by single power-law with a decay index of $1.03 \pm 0.06$ \citep{Evans_2009} and this gives the minimum jet break time to be 1 day.
The redshift of the GRB was established spectroscopically to be $z = 2.006$ \citep{Tanvir_2013}.

\subsection{GRB 140430A}
\citet{kopac_2015} presented polarization and photometric results for GRB 140430A. Here we performed a simple power-law fit as we did for other GRBs for consistency. For RINGO3 data, we find a decay index of $\alpha = 0.71 \pm 0.06, 0.57 \pm 0.02,$ and $0.55 \pm 0.02$ for $I$, $R$ and $V$ filters respectively (Fig~\ref{fig:photometry_1}). We note that here we are using 60 second stacked data whereas \cite{kopac_2015} used 10 second exposure data, hence there is difference in the light curve and decay index values. Swift XRT data was best fitted with three breaks at $320^{+17}_{-16}$, $412^{+26}_{-15}$, and $3.4^{+3.6}_{-1.6} \times 10^4$,seconds with a decay indices of $3.64^{+0.23}_{-0.19}, 8^{+0.0}_{-1.46},$, $0.64^{+0.06}_{-0.08}$, and $1.14 ^{+0.29}_{-0.23}$ \citep{Evans_2009}. The minimum jet break time is assumed to be 1.15 days from \citet{kopac_2015}. This burst is a relatively long burst with $T_{90} = 174 \pm 4$ and the X-ray light curve shows early flares which have been suggested to originate due to internal dissipation processes \citep{ Zhang_2006,Troja_2015, kopac_2015}. We performed a permutation analysis on the polarimetric data and found no probable detection. For the stacked data we find polarization upper limits of $<12.6\%$, $<9.6\%$, and $<16.5\%$ for $I$, $R$, and $V$ filters respectively. We find a similar upper limit of polarization as \citet{kopac_2015} for all the different wavelengths. We note that our upper limit values, as stated in Table~\ref{tab:polarization_nondetection}, are slightly different because we implement a different technique for error calculations and the time intervals of these measurements are different. The GISP estimates for this GRB are $1.03\%$, $1.17\%$, and $1.03\%$ for $I$, $R$ and $V$ band respectively. All the results presented in Fig.~\ref{fig:pol_results_1} are for 10 minute stacked data.

\subsection{GRB 141220A}
RINGO3 made observations of the GRB about 3 minutes after the trigger time of 6:02:52 UTC in all the three USNO $I$, $R$, and $V$ bands. IO:O SDSS $r'$ band observations were made 33 minutes after the trigger. We used field stars to calibrate magnitude and flux. Galactic extinctions of $A_V =0.035 $, $A_R =0.027$, $A_I = 0.02 $, $A_{r'} =0.029$ were corrected. The redshift of 1.3195 was inferred from spectroscopic observations done using OSIRIS at the 10.4 m GTC \citep{postigo_2014} and gamma-ray burst duration is $T_{90} = 7 \pm 0.5 s$ \citep{Lien_2016}. We fit a power-law function to RINGO3 observations of GRB 141220A and get a decay index of $\alpha$ = $1.09 \pm 0.02, 1.10 \pm 0.02$, and $1.03 \pm 0.02$ for $I$, $R$ and $V$ filters respectively as shown in Fig.~\ref{fig:photometry_1} and these values match well with the decay indices reported in \citet{Jordana_2021} of $1.105\pm 0.013$, $ 1.067 \pm 0.009$, and $1.095\pm 0.005$ for $I$, $R$, and $V$ filters. The Swift XRT light-curve could be fitted by broken power-law with a time break at $207^{+101}_{-45}$\,s and decay index of $-0.3 \pm 0.6$ and $1.375^{+0.104}_{-0.099}$ before and after the break \citep{Evans_2009}. \citet{Jordana_2021} reports a jet break time of 0.35 days or longer. 

 We present upper limit on bias-corrected polarization degree in Fig.~\ref{fig:pol_results_1} for all three wavelengths for 10 minute stacked data. \citet{Jordana_2021} found polarization detection for the first epoch in $V$ band and upper limits for the rest. However, in our analysis we do not find any detection and only upper limits for all the cases. This could be due to the difference in time bin of the presented results. Though the upper limit values from this analysis doesn't match exactly with results from \citet{Jordana_2021}, the trend of polarization for different filters i.e. $V$, $R$, and $I$ bands have an upper limit in incremental order is the same. In addition, the behaviour with time is consistent for all the filters with results from \citet{Jordana_2021}. The observed polarization degree data for the 30 minutes stacked data are $< 3.52\%, <3.23\%$, and $1.74\%$ for $I$, $R$ and $V$ bands respectively. For this GRB, the GISP is estimated to be $0.07\%, 0.08\%$, and $0.09\%$ for $I$, $R$, and $V$ bands thus contribution from GISP is negligible. Since we are observing forward shock dominated emission, the low level of polarization detection is in line with theoretical predictions \citep{Rossi_2004,Kobayashi_2019}. 

\subsection{GRB 151215A}

LT observations started within 3 minutes of the trigger time 3:01:28 UTC. Spectroscopic analysis of NOT observations gave the redshift of $z = 2.59$ \citep{Xu_2015}. The gamma-ray burst duration is $T_{90} = 18 \pm 1$\,s \citep{Gibson_2015}. We fit a single power-law and get $\alpha = 0.68 \pm 0.03, 0.98 \pm 0.03,$ and $0.92 \pm 0.03$ for $I$, $R$ and $V$ filters respectively. In Fig.~\ref{fig:photometry_1}, the decay index is slightly different for the $I$ band compared to the $R$ and $V$ bands. We note that there is another light source close to the target which could contaminate the GRB's aperture photometry in some cases. Thus, we cannot confirm colour evolution of the GRB. Swift XRT data is best fitted by a power-law of decay index $0.95^{+0.06}_{-0.05}$ \citep{Evans_2009}and this data put a lower limit on the jet break time to be 2.3 days.

We present bias-corrected polarization degree values at different times for GRB 151215A in Fig.~\ref{fig:pol_results_1}. We do not detect polarization and upper limit of observed polarization is presented. The polarization values for 40 minutes of stacked data are $< 6.45\%, <6.64\%$, and $6.05\%$ for $I$, $R$ and $V$ bands respectively. The estimated GISP values are $2.95\%, 3.35\%$, and $3.6\%$ for $I$, $R$, and $V$ bands respectively, which is a significant factor compared to the upper limit values.

\subsection{GRB 180325A}
RINGO3 observations started $\sim$13 minutes from the trigger time 01:53:02 and reliably detected the transient in the $I$ filter only. In Fig.~\ref{fig:photometry_2} we present the $I$ band light curve of the GRB along with limited, late-time IO:O $r'$ data.  We find a decay index of $\alpha = 0.58 \pm 0.04$ which is less than 1; we attribute it to the GRB forward shock for RINGO3 I data with possibility of energy injection.
Swift constrained the gamma-ray burst duration to $T_{90}=94 \pm 2 s$ \citep{troja_2018}. The redshift was obtained spectroscopically by NOT as $z = 2.25$ \citep{Heintz_2018}. The Swift XRT best-fitting light-curve has 3 breaks at $238^{+31}_{-116}, 2128^{+812}_{-461}, 3.5^{+0.6}_{-0.4}\times 10^4$ seconds with decay indices of $-0.75^{+0.22}_{-0.68},0.24^{+0.27}_{-0.25},1.99\pm 0.08,$ and $5^{+3}_{-2}$ \citep{Evans_2009}. Using this XRT light-curve, we assume a lower limit on jet break time to be 0.4 days. 

 The polarization degree values after bias correction are presented in Fig.~\ref{fig:pol_results_1} for $I$ band. Permutation analysis on the polarization values did not show any significance for all the observed data points. For the first 30 minutes of stacked data we obtain $< 12.29\%$ for $I$ band. The GISP for this case is $0.15\%$ in $I$ band.

\subsection{GRB 180618A}
RINGO3 made observations of the GRB about 3 minutes after the trigger time of 0:43:13 UTC in $I$, $R$, and $V$ bands. After 30 minutes of observations by RINGO3, IO:O was online and made follow-up observations of the GRB in SDSS $r$ band filter. Stars in the field of view were used to calibrate the magnitude and flux of the GRB and Galactic extinction of $A_V =0.182 $, $A_R =0.144$, $A_I = 0.103 $, $A_{r'} =0.155$ correction was implemented in the results. The gamma-ray duration is $T_{90} = 3.7 \pm 0.6 $ s from Fermi GBM observations and Swift UVOT filter detection put the upper limit on the redshift to be $z<1.2$ \citep{GCN22810}.

The light curve from Swift-BAT data shows a short multi-peak at $T_0$ to $\sim T_0+0.3 s$ and extended emissions lasting until $\sim T_0+50 s$ \citep{Sakamoto_2018}. They also did further analysis to get power-law index and fluence which are consistent with a short GRB with extended emission \citep{Sakamoto_2018}. The Swift XRT light curve is best fitted by a power-law with 3 breaks at $147^{+21}_{-22}, 296^{+185}_{-55},$ and $5483^{+1.63\times 10^3}_{-2015}$ s with decay indices of $0.80^{+0.17}_{-0.20}, 1.48\pm 0.19, 1.876^{+0.132}_{-0.063},$ and $1.04^{+0.18}_{-0.15}$ \citep{Evans_2009}.

The best fit for the RINGO3 data at all three different wavelengths is a broken power law with the same break time of $1370$ s. We assume this break time to be the lower limit of jet break time as well. The best fit has  $\alpha = 0.48 \pm 0.08, 0.53 \pm 0.04,$ and $0.57 \pm 0.05$ for $I$, $R$ and $V$ filters respectively before the break and $\alpha = 2.32 \pm 0.8, 2.45 \pm 0.4,$ and $2.26 \pm 0.5$ for $I$, $R$ and $V$ filters respectively after the break.

GRB 180618A is a short GRB with extended emission as discussed by \cite{Sakamoto_2018}. Our upper limits on polarization are large for $I$ and $R$ filters at a later time because the source is fainter and the noise is high as shown in Fig.~\ref{fig:pol_results_2}. For the $V$ filter the upper limit values are better constrained because of the higher signal-to-noise ratio in this filter for the source. For 30 minutes of stacked data we get polarization values of $< 5.26\%, <12.24\%$, and $<6.78\%$ and GISP values of $0.51\%, 0.58\%$, and $0.63\%$ for $I$, $R$, and $V$ bands respectively. Further detailed analysis of GRB 180618A RINGO3 data will be presented by Jordana-Mitjans et al., 2022 (submitted).

\subsection{GRB 190114C}
The LT observed this GRB $\sim$ 3 minutes after the burst time of 20:57:02.341 UTC and made observations using RINGO3 in $I$, $R$, and $V$ bands. After the first thirty minutes of RINGO3 observations, IO:O was triggered and made observations in SDSS $r$ band. Since the GRB was bright, more RINGO3 observations were taken after the IO:O observations. Detailed analysis of the LT follow-up observations and data from other telescopes has been presented in \citep{Jordana_2020}. Here we present a similar analysis to other GRBs. The light curve and power-law fit agree well with the results from \citet{Jordana_2020} as seen in Fig.~\ref{fig:photometry_2}. The best-fit for the RINGO3 light curve is a broken power-law with a break at $401$ seconds (we find the best fit to have the same break time for all the filters unlike in \citet{Jordana_2020}). We get $\alpha = 1.43 \pm 0.03, 1.50 \pm 0.02,$ and $1.47 \pm 0.02$ for $I$, $R$ and $V$ filters respectively before the break and $\alpha = 0.87 \pm 0.02, 0.94 \pm 0.01,$ and $0.99 \pm 0.02$ for $I$, $R$ and $V$ filters respectively after the break. The jet break time is 0.21 days \citep{Jordana_2020}.

\citet{Jordana_2020} have presented detailed polarimetric analysis of GRB 190114C using RINGO3 data. Here we perform our polarimetric RINGO3 GRB analysis for 10 minute stacked data. We obtain a polarization detection for the earlier time period and the polarization values are low; mostly coming from the ISP of the host galaxy as seen by \citet{Jordana_2020}. We note a slight difference in polarization measurements compared to \citet{Jordana_2020} due to a difference in time intervals of our measurements and a different error calculation technique. For these data points we performed permutation analysis and found detections for a few points as presented in Table~\ref{tab:polarization_detection}. There are two points in the $I$ band for the time interval 803-1403 and 1403-2003 seconds and one data point in the $R$ band for the time interval 803-1403 seconds whose error values do not cross the zero point, however, their permutations ranks are lower than 0.95. Hence, we do not consider these values as detection and present them as the upper limit in Fig.~\ref{fig:pol_results_2}. For 30 minutes of stacked data we get polarization values of $< 2.65\%, <2.78\%$, and $<2.21\%$ and GISP values of $0.07\%, 0.08\%$, and $0.09\%$ for $I$, $R$, and $V$ bands respectively. \cite{Jordana_2020} estimated the polarization contribution of the host galaxy to be $< 3.9\%$, $< 4.5\%$, and $< 4.5\%$ (larger than the detected polarization) therefore the detected polarization could easily be interpreted as simply coming from the dust in the host galaxy confirming our earlier work \citep{Jordana_2020}. 

\subsection{GRB 191016A}
There was a delay in LT observations of this GRB and initial IO:O observations were made in SDSS $r$ band 40 minutes after the trigger time of 04:09:00 UTC. We made RINGO3 follow-up observations 66 minutes after the trigger time in $I$, $R$, and $V$ filter. Even though it was observed much later, the afterglow was bright enough to be detected in all the filters. Magnitude and flux calibrations were done with stars in the field and Galactic extinction of $A_V =0.281 $, $A_R =0.222$, $A_I = 0.159 $, $A_{r'} =0.239$ was also corrected. The gamma-ray burst duration was inferred to be  $T_{90} = 220 \pm 183$ seconds from Swift and the photometric redshift of the burst is 3.29 $\pm$ 0.40 \citep{Smith_2021}.

Detailed analysis for the GRB is presented in \citet{Shrestha_2021}. Briefly, the light curve is best fitted by broken power-law with different break point for different filters. For initial IO:O data the best fit model shows a simple power-law decay with decay index of 1.24. For RINGO3 $I$, $R$, and $V$ bands the best fit models have decay index of $0.97 \pm 0.07, 0.98 \pm 0.07,$ and $1.25 \pm 0.1$ before the break time of 6146, 6087, and 5247 seconds. After the break the decay indices are $0.04 \pm 0.17, -0.44 \pm 0.17,$ and $0.01 \pm 0.09$ respectively. This plateau phase is also seen by \citet{Pereyra_2022}. The difference in decay indices during the plateau phase is hard to explain. One possibility presented by \citet{Shrestha_2021} is the difference in electron energy distribution indexes in the blast wave and the reverse shock because the reverse shock is sub-relativistic. Further analysis could be found in \citep{Shrestha_2021}. \citet{Shrestha_2021} calculated the jet break time for this GRB to be 0.53 days with a limited number of data points. Later, \citet{Pereyra_2022} calculated the jet break time using a larger number of data and found it to be between 0.24 to 0.52 days after the trigger. Using these two values along with Swift XRT data, we present 0.52 days as jet break time for this GRB.

We present results of bias corrected polarization degree in Fig.~\ref{fig:pol_results_2}. We get polarization detection at 1 sigma level in all three filters at different time period. One hour stacked data shows polarization values of $<3.83\%$, $<5.23\%$, and $<3.7\%$ respectively. The GISP estimates for this GRB are $0.66\%$, $0.75\%$, and $0.8\%$ in $I$, $R$, and $V$ bands, thus a very negligible contribution to our polarization measurements. We also calculated the ISP contribution of the host galaxy for this GRB. The best fit model from \citet{Smith_2021} shows $A_V =0.354$ for the host galaxy, and calculating the ISP using this extinction value for Milky Way-like dust gives $1.0\%$, $1.1\%$, and $1.2\%$ for $I$, $R$, and $V$ bands. However, in \citet{Smith_2021} the best fit model shows small magellanic cloud (SMC)-like dust for the host galaxy and using SMC-like dust \citep{Rodrigues_1992} to estimate host galaxy ISP gives $0.92\%$, $1.18\%$, and $1.45\%$ for $I$, $R$, and $V$ bands respectively. Hence, the detected polarization is intrinsic polarization.

Our polarization result matches well with our previous analysis presented in \citet{Shrestha_2021} which showed that the combination of polarimetry and photometry favours scenarios with energy injection from the central engine.  In this case, slower magnetized ejecta from the central engine catches up with the decelerating blast wave and causes forward and reverse shocks. This short-lived reverse shock can explain the polarization detection we see near the plateau phase of the light-curve.

\section{Discussions} \label{sec:discussions}

We have presented polarimetric and photometric analysis of ten GRBs observed by RINGO3.
Four of them, GRB 140430A, GRB 141214A, GRB 190114C, and GRB 191016A have been published separately in \citet{kopac_2015,Jordana_2020, Jordana_2021, Shrestha_2021}, in this paper we have carried out a uniform re-reduction and analysis of the whole sample. Our analysis produces similar values of polarization degree and EVPA for these GRBs as previously noted in the published papers. We have presented light curves of GRB 130606A, 130610A, and 130612A which have been fit with a power-law; we have presented the decay indices ($\alpha$) for these GRBs for all the three RINGO3 wave bands except for GRB 130606A for which we only have good SNR for the $I$ band as shown in Fig~\ref{fig:photometry_1}. The XRT light-curve for GRBs in the same time period is also presented. For all sources, other than GRB 140430A, we see the XRT light curve is similar to their optical light curves. Hence the X-ray photons in these events should originate from the forward shocks.

For most of the cases, the decay indices for the photometric light curve are less than 1.5 which shows we are observing the forward shock dominated light-curve \citep{Sari_1999, Kobayashi_2000, Zhang_2004, Gomboc_2009, Japelj_2014} in some cases with energy injection (decay indices can be closer to 0.5). However, for the case of GRB 130606A and GRB 180618A (after the break) we observed decay index values greater than 1.5. For GRB 130606A it could be the reverse shock dominated emission we are seeing but we do not have polarization degree calculations due to the instrument not being well calibrated. And for GRB 180618A, the steeper decay could be due to the jet break instead of a reverse shock emission. However, we note that the flattening in the X-ray at later times (alpha = 1.04 at $t>5483s$) can not be 
explained in a simple jet model. GRB 151215A and  GRB 180325A have light curves which could be modelled by a single power-law. Their decay indices are smaller than 1.5; suggesting that the forward shock emission is dominating or suggestive of some energy injection. For these cases, we get polarization upper limits in three wavelengths, which is expected in the case of forward shock dominated emission \citep{Rossi_2004}. Thus, it is possible that most of the observed polarization is contributed by dust in the host galaxy. GRB 180618A is a short GRB with extended emission. The light-curve of this GRB is best fit by a broken power-law with a break at $1370$ seconds. Initially the light-curve showed a shallow decay of $0.48$, $0.53$, and $0.57$ for $I$, $R$, and $V$ band respectively. After the break the decay is sharper with $2.32$, $2.45$, and $2.26$ for $I$, $R$, and $V$ band respectively. There are few points after the break, thus the sharp decay is not well modelled. We do not detect any polarization and get upper limits in polarization. Further discussion on GRB 180618A will be presented in Jordana-Mitjans et al. 2022 (submitted). 

\begin{figure*}
    \centering
    \includegraphics[width=2\columnwidth]{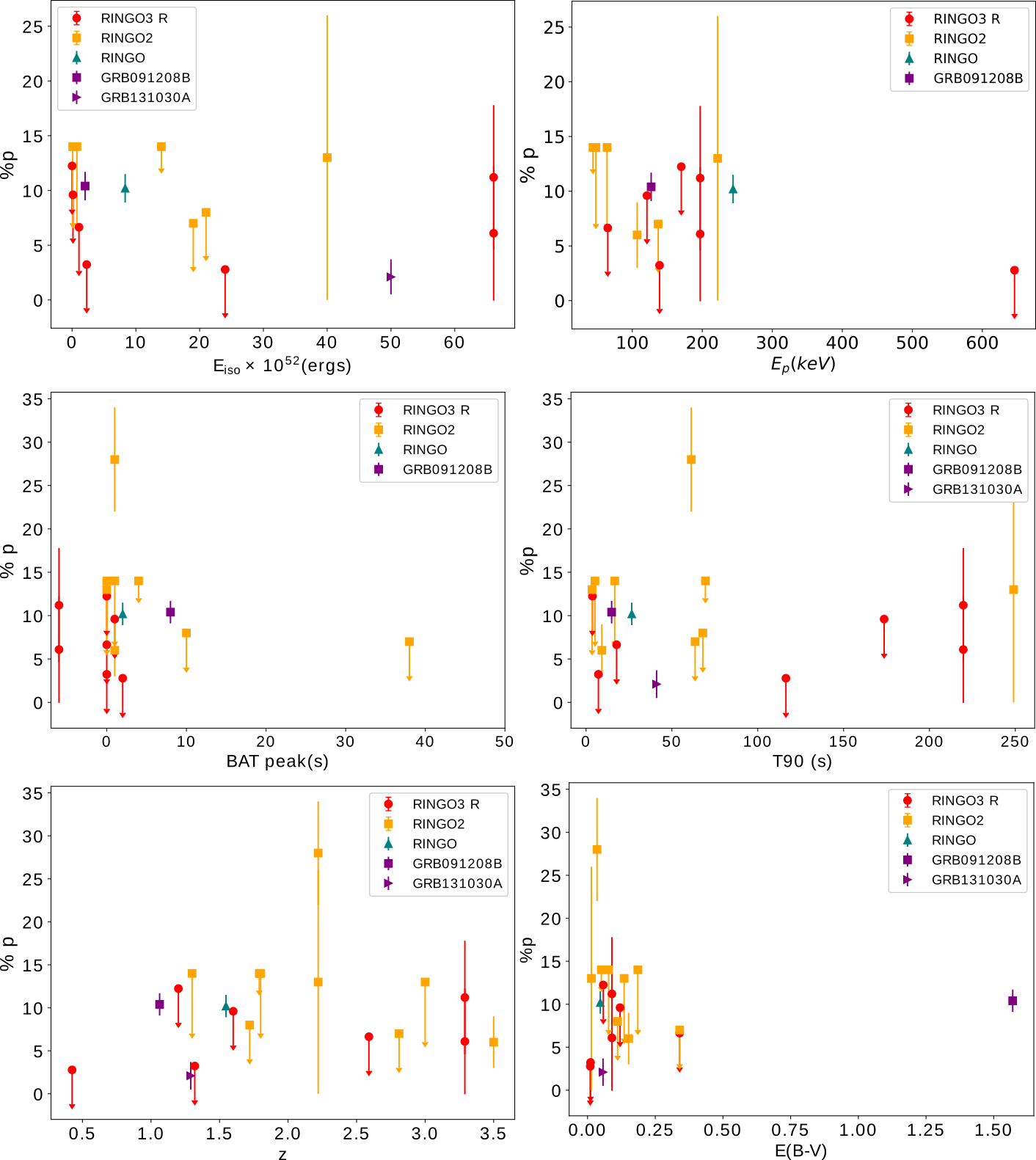}
    \caption{Observed polarization values (stacked) from RINGO3 along with all the early time optical afterglow polarization observations in the literature with respect to different GRB properties. Upper limit are presented as downward arrow. Polarization values for RINGO2 are from \citet{steele-ringo2-2017}, for RINGO are from \citet{Steele_2009}, GRB 091208B is from \citet{Uehara_2012}, and GRB 131030A is from \citet{King_2014}. }
    \label{fig:pol-properties}
\end{figure*}

There are few polarization observations of GRB early afterglows in the literature. \citet{Uehara_2012} detected polarization in the early afterglow of GRB 091208B and \citet{King_2014} reported early-time polarization of GRB 131030A. We investigated the relationship between the polarization signal and various properties of GRBs such as decay index in Fig.~\ref{fig:alpha}, isotropic energy $E_{iso}$, peak energy $E_p$, BAT peak, $T_{90}$, redshift, and extinction of the Milky Way as shown in Fig.~\ref{fig:pol-properties}. In addition, we also checked the relation between polarization and temporal distance of jet break from our observation. As most of the data points are upper limits, we performed survival analysis \citep{Feigelson_1985} using the Python package lifelines \citep{lifeline_2020} to check for any co-relation between polarization signal and different GRB parameters as noted. For all of the cases we get a concordance index close to 0.5, which is the expected results from random predictions, hence, we cannot conclude any relation from our data set. In order to get a better relation between polarization and various properties we need to increase the number of observations of GRB early afterglows. The increased sensitivity of the new polarimeter MOPTOP \citep{Shrestha_2020} on the LT will improve the number of polarization observations of early afterglows in the future.

\subsection{Polarization and decay index}

\begin{figure*}
    \centering
    \includegraphics[width=2\columnwidth]{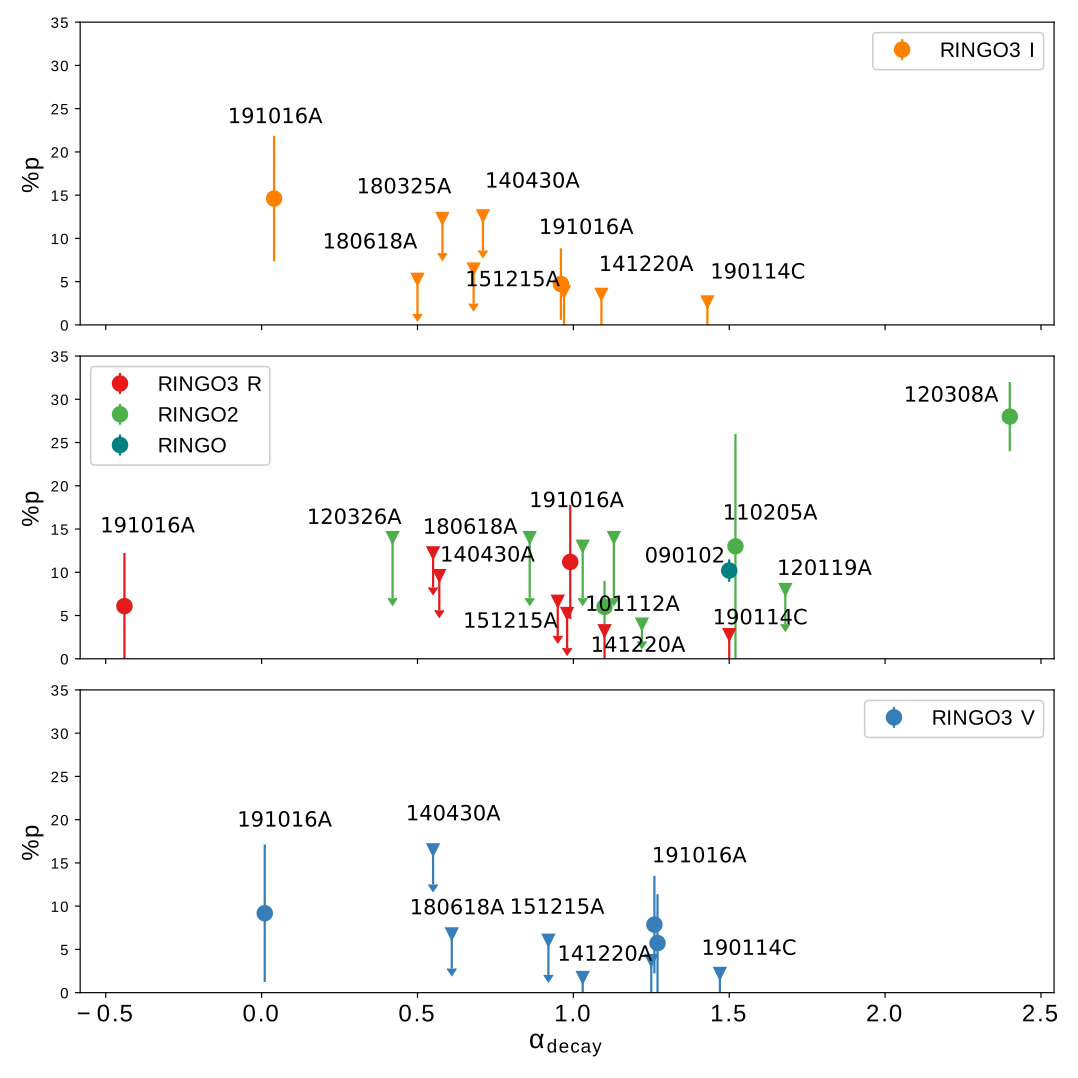}
    \caption{Observed polarization values (all observed data stacked for upper limit along with detected polarization values) with respect to the decay indices deduced from the light curves. Three panels show results for $I$, $R$, and $V$ band results from top to bottom respectively. For $R$ band, we also present results from RINGO2 observations previously presented in \citet{steele-ringo2-2017}. Names of the GRBs are given for all the cases except for GRB 100805A, GRB 110726A, GRB 120311A, and GRB 120327A from RINGO2 observations for better visibility of points in the plot. Polarized detection is presented as a circle and upper-limit is presented as a inverted triangle. For GRB 191016A, we present all the detected polarization for the different time periods and the decay index value of the corresponding time period is plotted.}
    \label{fig:alpha}
\end{figure*}

In the literature we find most of the high degree of polarization values to be observed for the case of reverse-shock emission \citep{Steele_2009, Mundell_2013}. When the GRB jet interacts with the local ambient medium, there are forward and reverse shock components: the reverse shock is short lived emission and decays faster than the forward shock emission. The decay index for the light-curves where reverse shock is dominating is expected to be greater than 1.5 \citep{Sari_1999, Kobayashi_2000, Zhang_2004, Gomboc_2009, Japelj_2014}. Previously, \citet{steele-ringo2-2017} presented how polarization varies with decay index for 9 different GRBs observed when RINGO2 was online (see Figure. 13 in their paper). Here we add to this data set and study how polarization changes with decay index for $I$, $R$, and $V$ bands in Fig.~\ref{fig:alpha} top, middle, and bottom panels respectively. In the middle panel, we also include results from RINGO2 observations as presented in \citet{steele-ringo2-2017}. All the upper limits presented here are for stacked data shown in Tables~\ref{tab:polarization_nondetection} and ~\ref{tab:polarization_detection} and B 190114C and GRB 191016A detected polarization is also presented. For GRB 191016A, polarization has been detected before the break in light-curve and after the break which is also plotted in Fig.~\ref{fig:alpha}. In the RINGO3 data set we do not find any cases with a decay-index close to 2, thus we suggest that our observed cases are for forward shock dominated emission, which is not highly polarized as shown in the Figure~\ref{fig:alpha}. To get a better relation between polarization and decay-index we need to increase the number of observations of GRB early afterglows, where the reverse shock is dominated which will be possible thanks to new polarimeter MOPTOP \citep{Shrestha_2020} on the LT.

\section{Conclusions} \label{sec:conclusions}
We have presented photometric and polarimetric results and analysis of ten GRBs out of 67 GRBs triggered by RINGO3 during the time period of 2013 to 2020. For the first three GRBs, instrument polarization was not well constrained so we only present photometry results. For the subsequent seven GRBs we present both photometric and polarimetric results with polarization degree (or upper limits) and EVPA values in the case of detection. Out of these GRBs, polarization was detected for GRB 190114C and GRB 191016A. Further analysis of GRB 190114C showed that detected polarization was contributed by the host galaxy dust. For GRB 191016A the contribution from the host galaxy, assuming SMC-like dust, is negligible and thus the detected polarization is considered to be intrinsic.

We created light-curves of all ten GRBs using RINGO3 for $I$, $R$, and $V$ band (where available), IO:O $r'$ band and RATCam $r'$ band data. We performed best fits for these RINGO3 light-curve using either single power-law or broken power-law and report the decay indices for these light curves. We analyzed the relation between decay index and polarization degree, since we mostly observed slowly decaying events we cannot provide clear correlation between decay index and polarization degree. We performed survival analysis to investigate if there is any co-relation between decay index and polarization. From our survival analysis we get concordance index of 0.47 which shows that with our limited data, we do not see any co-relation. Hence, we need more early time observations of GRB events to study the relation between polarization and decay index.

We make an intrinsic detection of polarization for GRB 191016A which has a late peak of at least 1000 seconds after the BAT trigger \citep{Smith_2021}. The source is bright enough to perform polarimetery and photometry even 66 minutes after the BAT trigger. The light curve is best fitted by a broken power-law at 5500 seconds after the BAT trigger and we get a shallow decay index close to 1 for all three wavelengths before the break time and it plateaus after the break time. With a high level of detected polarization ($>9\%$) and no jet-break like feature, we deduce that the light-curve has reverse shock emission and the shallow decay is due to the energy injection to the forward shock/ blast wave.  

The GRB 190114C case shows that even a detection of low polarization degree can help us understand the afterglow emission mechanism. For the GRB 191016A case, polarization and EVPA calculations along with the light-curve allowed us to carry out detailed analysis of the afterglow emission. In the absence of polarization analysis, the GRB 191016A afterglow would have been considered as forward shock emission. However, polarization and EVPA measurements point towards the possibility of reverse shock emission in the afterglow. Thus, polarization observations of GRBs can provide crucial clues to getting detailed information about the event along with photometric and spectroscopic observations.

 RINGO3 polarimeter have successfully observed various early optical afterglows of GRBs within few hundred seconds of trigger. The results presented in this paper shows the importance of simultaneous multicolour photometry and polarimetry (colours>2) which helps to determine the underlying emission mechanism. New polarimeters with increased sensitivity to probe a larger statistical sample over a significant time period of their evolving emission would open new windows on GRB physics.

\section*{Acknowledgements}
Operation of LT on the island of La Palma by Liverpool John Moores University at the Spanish Observatorio del Roque de los Muchachos of the Instituto de Astrofisica de Canarias is financially supported by the UK Science and Technologies Facilities Council (STFC). Financial support for the development of MOPTOP was provided by the STFC PRD scheme. MS is supported by an STFC consolidated grant number (ST/R000484/1) to LJMU. We acknowledge with thanks the variable star observations from the AAVSO International Database contributed by observers worldwide and used in this research. This research made use of Photutils, an Astropy package for detection and photometry of astronomical sources (\cite{Bradley_2019}). AG acknowledges the financial support from the Slovenian Research Agency (research core funding P1-0031, infrastructure program I0-0033, project grants J1-8136, J1-2460) and networking support by the COST Actions CA16104 GWverse and CA16214 PHAROS. CGM and NJM thank Hiroko and Jim Sherwin for financial support. This work made use of data supplied by the UK Swift Science Data Centre at the University of Leicester. We would like to thank our anonymous referee for thorough and thoughtful comments that have greatly improved the paper.

\section*{Data Availability}
All the observational data are freely available online in the LT archive at https://telescope.livjm.ac.uk/.


\bibliographystyle{mnras}
\bibliography{grb_catalog} 








\bsp	
\label{lastpage}
\end{document}